\documentclass[]{aastex}
\usepackage{graphicx}
\usepackage{emulateapj5}

\begin{document}

\slugcomment{Accepted for publication in PASP, Vol 114, December 2002}

\title{A Detailed Thermal Analysis of the Binospec Spectrograph}

\author{Warren R.\ Brown, Daniel G.\ Fabricant, and David A.\ Boyd}
	\affil{Harvard-Smithsonian Center for Astrophysics, 60 Garden
Street, Cambridge, MA 02138}
\email{wbrown@cfa.harvard.edu,dfabricant@cfa.harvard.edu,dboyd@cfa.harvard.edu}

\shorttitle{Spectrograph Thermal Analysis}
\shortauthors{Brown, Fabricant,\& Boyd}

\begin{abstract}

	Refractive optics in astronomical instruments are potentially
sensitive to temperature gradients and temperature transients.  This
sensitivity arises from thermally dependent refractive indices, lens
spacings, and lens dimensions.  In addition, thermal gradients in the
instrument structure can cause undesirable image shifts at the
detector that degrade instrument calibration.  We have therefore
undertaken a detailed thermal analysis of Binospec, a wide-field
optical spectrograph under development for the converted MMT.  Our
goals are to predict the temperature gradients that will be present
in the Binospec optics and structure under realistic operating
conditions and to determine how design choices affect these
gradients.  We begin our analysis by deriving thermal time constants
for instrument subassemblies to estimate the magnitude of temperature
gradients in the instrument and to determine where detailed thermal
models are required.  We then generate a low-resolution finite
difference model of the entire instrument and high-resolution models
of sensitive subassemblies.  This approach to thermal analysis is
applicable to a variety of other instruments.

	We use measurements of the ambient temperature in the converted
MMT's dome to model Binospec's thermal environment.  In moderate
conditions the external temperature changes by up to 8 $^{\circ}$C over
48 hours, while in extreme conditions the external temperature changes
by up to 17 $^{\circ}$C in 24 hours.  During moderate conditions we
find that the Binospec lens groups develop $\sim$0.14 $^\circ$C axial
and radial temperature gradients and that lens groups of different mass
develop $\sim$0.5 $^\circ$C temperature differences; these numbers are
doubled for the extreme conditions. Internal heat sources do not
significantly affect these results; heat flow from the environment
dominates.  The instrument must be periodically opened to insert new
aperture masks, but we find that the resulting temperature gradients
and thermal stresses in the optics are small.  Image shifts at the
detector caused by thermal deflections of the Binospec optical bench
structure are $\sim$0.1 pixel hr$^{-1}$. We conclude that the proposed
Binospec design has acceptable thermal properties, and briefly discuss
design changes to further reduce temperature gradients.

\end{abstract}


\section{INTRODUCTION}

	The refractive optics used in modern astronomical
instruments, including multi-object spectrographs and wide-field focal
reducers, must typically be designed to operate over a wide range of
temperatures.  Refractive indices, lens spacings and lens dimensions
are all temperature dependent, and athermal designs are necessary to
compensate for changes in these quantities.  However, it is difficult
or impossible for the optical designer to compensate for the
temperature gradients arising from temperature transients.

	In this paper we study the time dependent thermal behavior of
Binospec, a wide-field, multi-aperture spectrograph being developed
for the 6.5 m converted MMT.  The mechanical layout of the Binospec
spectrograph is shown in Figures \ref{fig:binotop} and
\ref{fig:binobot}.  Binospec uses an ambitious refractive focal
reducer (collimator and camera) \citep{fabricant98, epps98} to image
two adjacent 8$\arcmin$ by 15$\arcmin$ fields.  Binospec operates at
the wide-field f/5 focus of the converted MMT \citep{fata93}.

	The collimators (see Figure \ref{fig:binocol}) each contain nine
elements in three groups and produce a 200 mm diameter collimated beam.
The cameras (see Figure \ref{fig:binocam}) each contain ten elements in
four groups that image onto a 4K by 4K CCD array.  These elements are
mounted in aluminum bezels with athermal, annular room-temperature
vulcanized (RTV) rubber bonds \citep{fata98}.  The bezels have mounting
bases that are attached to Binospec's optical bench.

	The optics are athermalized using a new technique described in
\cite{epps02}. Briefly, the weak lenses formed in the fluid used to
couple the multiplets exploit the thermally sensitive refractive index
of the coupling fluid to compensate for thermally dependent lens
properties:  refractive indices, spacings, and dimensions.  However,
our concern that nonequilibrium conditions might degrade the
performance of these optics motivates the present thermal analysis.

	Our thermal analysis focuses on calculating the temperature
gradients in the optics and the time scale of temperature changes in
the instrument due to temperature transients arising from: (1) external
conditions in the telescope dome, (2) the intermittent powering of
internal heat sources (motors or other actuators) and (3) opening the
instrument to change aperture masks, filters, or gratings.  We also
calculate the thermal stresses in the optics and the thermal deflection
of the instrument optical bench. Our primary goals are to verify that
Binospec will maintain its specified image quality, stability, and
lateral scale under realistic operating conditions and to determine
what design modifications would be beneficial.

\includegraphics*[width=3.25in]{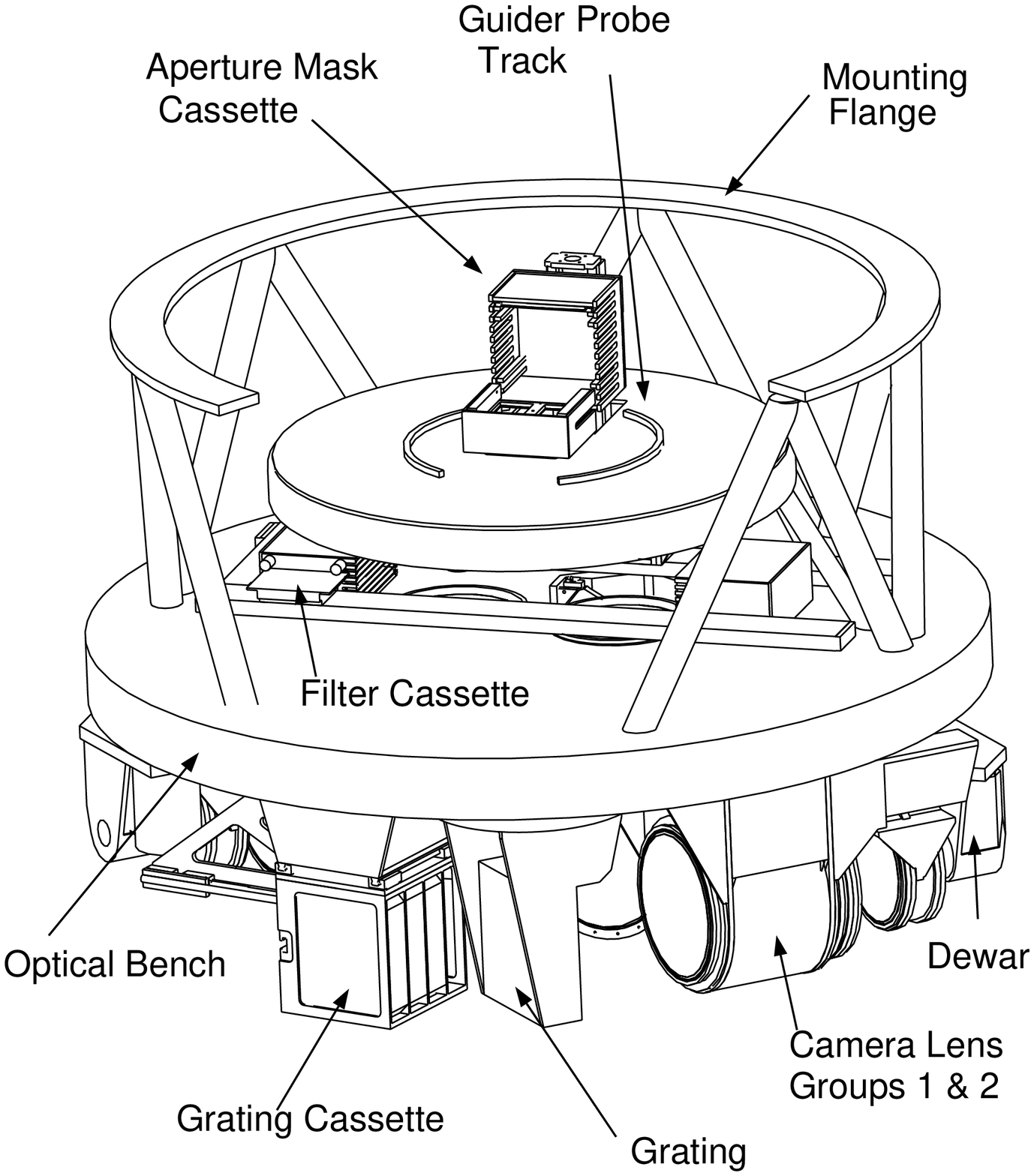}
	\figcaption{Binospec mechanical layout: top view.\\
\label{fig:binotop}}


	Thermal analyses usually follow one of two general
approaches.  One approach is to find a closed form analytical
solution for the heat transfer equations.  Binospec is too complex
for this approach.  The second approach is to create a detailed
thermal model of the entire instrument.  We find that a complete,
highly detailed model is not required to understand Binospec's
thermal properties.  Our approach to thermal analysis relies on a
mixture of coarse and detailed models, and we believe that this
approach is applicable to a variety of other instruments.

	Our paper is organized to follow the steps in the analysis.  
We begin in \S 2 by laying out the basic issues of heat transfer, and
in \S 3 summarize the quantities needed for the heat transfer
calculations. In \S 4, we begin the thermal analysis with calculations
of thermal time constants and thermal capacitances.  These analytic
calculations help us to understand how quickly various Binospec
components respond to temperature changes, allowing us to identify the
components that dominate the heat transfer.  We make detailed models of
these components and coarser models of less sensitive components. The
analytic calculations also provide an important check of our final
thermal models.  In \S 5 we create thermal finite difference models of
the instrument and its subassemblies to calculate heat flows and
temperature gradients.  We use a low-resolution model of the entire
spectrograph to provide boundary conditions for high-resolution models
of the subassemblies.  We model in detail the thermally sensitive
components, such as the collimator and camera optics, the optical
bench, and the filter changer. In \S 6 we discuss the results from
these models.  We conclude in \S 7.

\includegraphics[width=3.25in]{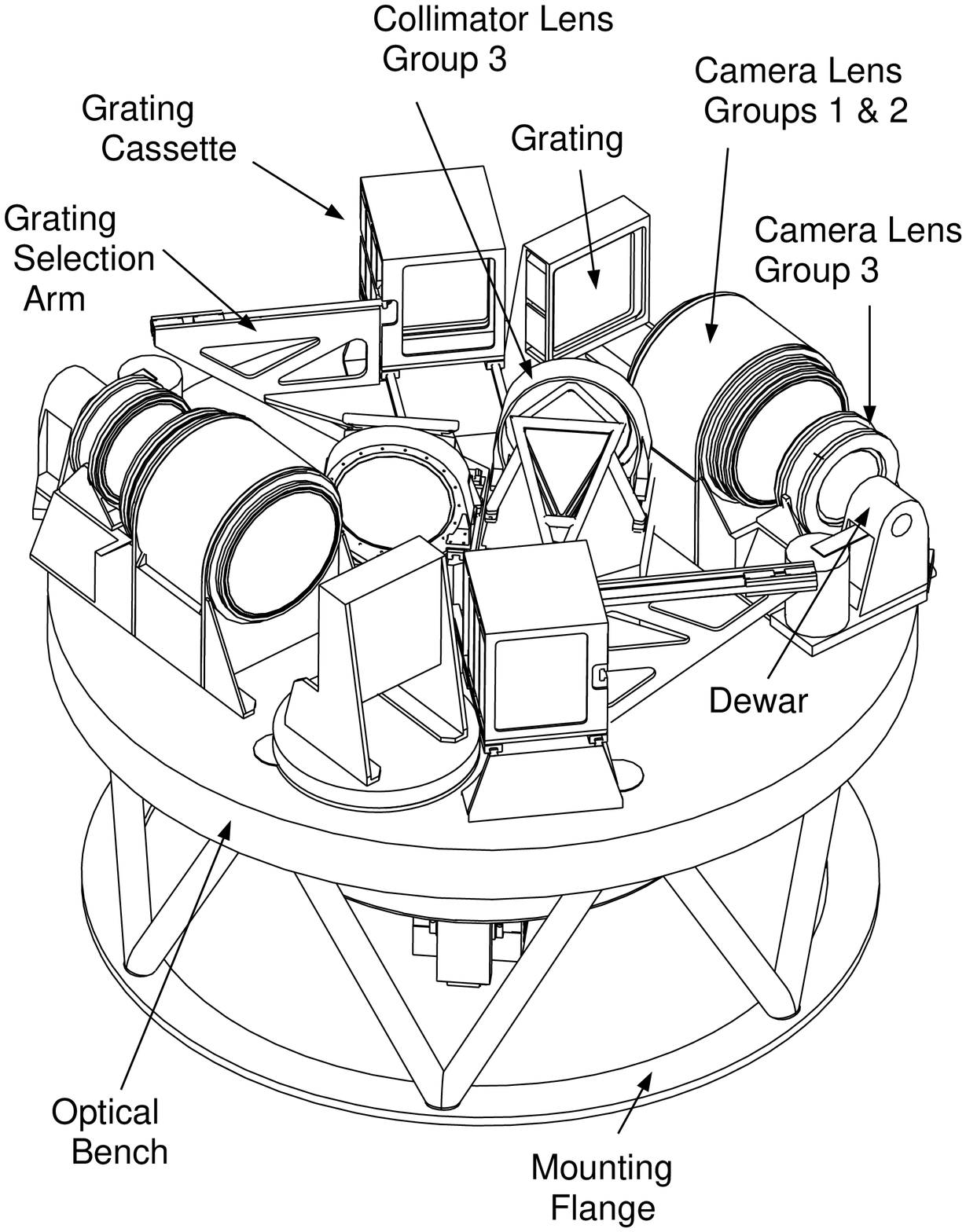}
	\figcaption{Binospec mechanical layout: bottom view. Collimator
lens groups 1 and 2 are hidden by the optical bench.\\
\label{fig:binobot}}

\vskip 80pt

\includegraphics[width=3.25in]{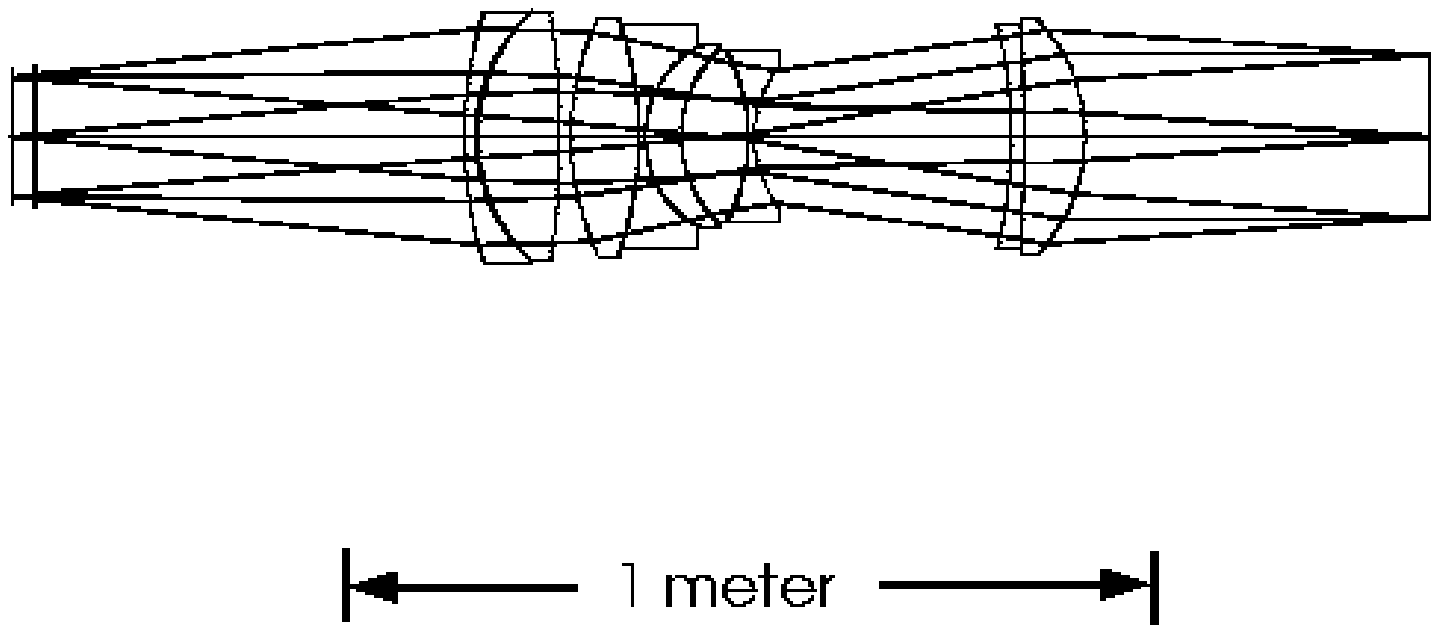}
	\figcaption{Optical layout of the Binospec collimator, with
fold mirrors removed.  From left to right the optical materials are:
BAL15Y, S-FSL5Y, PBM2Y, PBL6Y, BAL35Y, CaF$_2$, PBL6Y, BSM51Y,
CaF$_2$.\\
\label{fig:binocol}}

\clearpage

\includegraphics[width=3.25in]{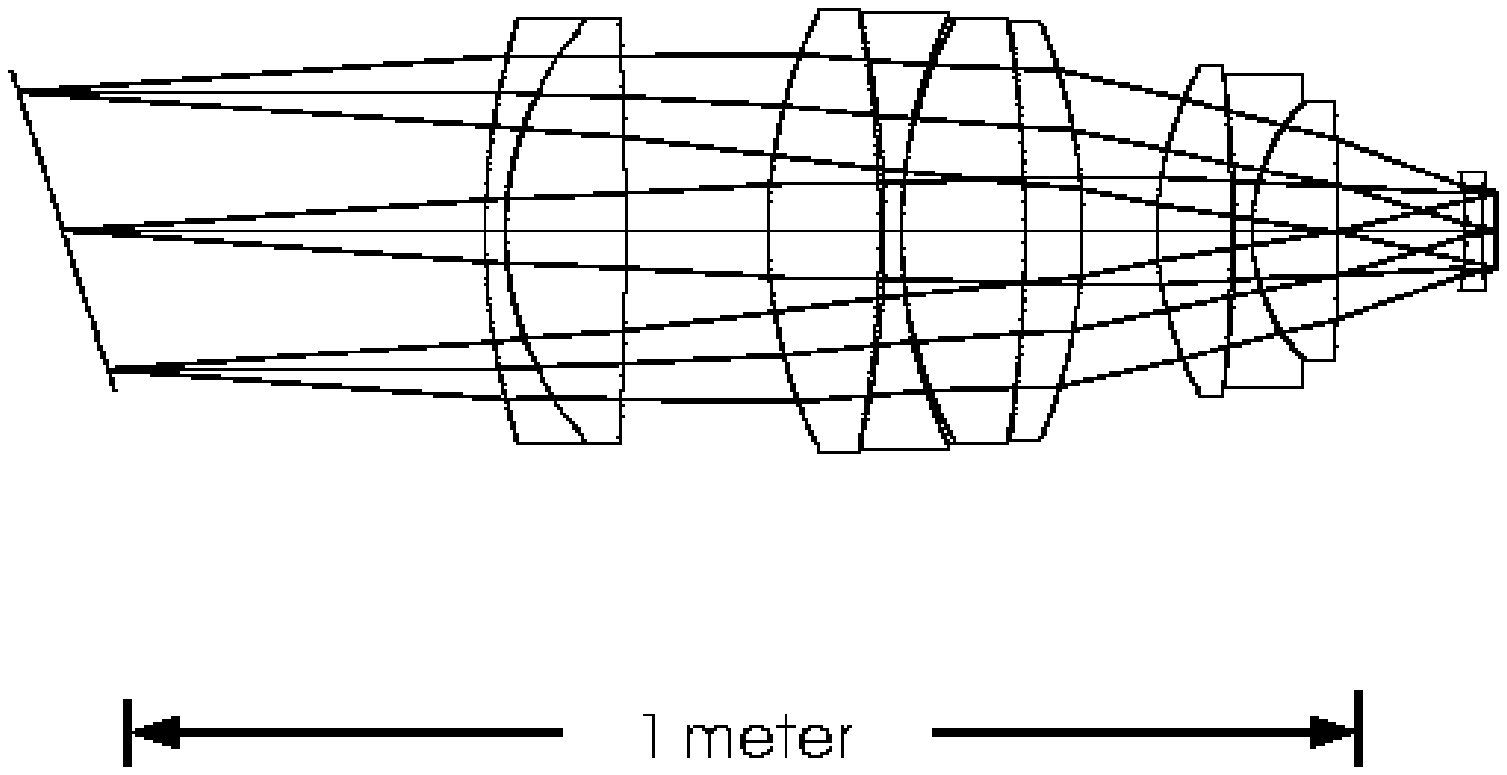}
	\figcaption{Optical layout of the Binospec camera.  The last
singlet serves as the dewar window. From left to right the optical
materials are: BAL35Y, CaF$_2$, CaF$_2$, BAL35Y, CaF$_2$, PBM2Y,
FPL51Y, NaCl, CaF$_2$, BSM51Y.\\
\label{fig:binocam}}

\section{HEAT TRANSFER}

Heat transfer is driven by temperature differences.  There are
three modes of heat transfer: conduction, convection, and radiation,
all of which are important for Binospec.  The heat transfer rates,
$Q$, for these three modes are:
	\begin{eqnarray}
	Q_{conduction} =& k A_x (T_1 - T_2) / L \label{condeqn} \\
	Q_{convection} =& h A (T_{air} - T_{surface})\\
	Q_{radiation} =& \epsilon \sigma A (T^4_{env} - T^4_{surface}),
	\label{radeqn}
	\end{eqnarray} where the variables are defined in Table
\ref{tab:symbols}.

Heat transfer is governed by the conservation of energy.  In a given
interval of time the heat flow into a point plus the heat generated at
that point minus the heat flow out equals the net energy gained or
lost, $q_{net}$. The net energy  will cause a change in
temperature scaled by the thermal capacitance, $mC$:
\begin{equation} q_{net} = m C (T_{final} - T_{initial}).
\label{eqn:mcdt}
	\end{equation}

	Thermal finite difference models divide a complicated system,
such as the Binospec spectrograph, into a number of lump masses that
are connected by the three modes of heat transfer.  Discrete time
steps are used to numerically solve the time dependent network of heat
transfer equations.  The time derivative of the temperature is
approximated by:
	\begin{equation} \label{time}
	\frac{dT}{dt} \approx \frac{T(t+\Delta t) - T(t)}{\Delta t}.
	\end{equation} Temperatures are calculated at times incremented by
$\Delta t$, so that $t = n \Delta t$.  The time step must be smaller
than the shortest time constant in the model; our time steps are
typically a few seconds.  We refer the reader to a heat transfer text,
such as \citet{incropera}, for more details.

\section{PREPARATION FOR THE THERMAL ANALYSIS}

The first step in the heat transfer analysis is to gather information
on the instrument's optical and mechanical designs, the materials
it uses, the environment in which it operates, and its intended
operational modes. We need to determine the mass, dimensions,
cross-sectional area, surface area, and surface emittance for each
instrument component.  The required level of detail depends on the
desired resolution of the results.  Temperature gradients in a
shutter, for example, will usually not affect the optical performance
of the spectrograph and so the shutter may be treated as a lump mass
with a single cross-sectional area and surface area.  Conversely,
temperature gradients in the optical bench are likely to be important,
and so its properties should be carefully detailed.  In addition we
need to determine the density, conductivity, specific heat, and
coefficient of thermal expansion for each material used in the
instrument.  The conductivities of steel and aluminum are orders of
magnitude larger than the conductivities of glass, leading to the
general result that instrument structures will equilibrate faster than
the optics they support.

The environment serves as the boundary condition for the
thermal model, and so we need to know the time dependence of the
temperature in the instrument operating environment.  We must also
determine if the instrument's surface is directly exposed to the night
sky or to a nearby heat source.  In addition, the air flow speed
affects the convection coefficient for the instrument's exterior
surface.  The amplitude and time scale of environmental temperature
variations turn out to be the dominant factors in determining the heat
flow and temperature gradients in Binospec.

The typical and extreme operating modes of an instrument should be
identified to determine the location, power output, and duty cycle of
internal heat sources.  Important internal heat sources include
motors, electronics boxes, and calibration lamps.

\vskip 25pt

\includegraphics*[width=3.25in]{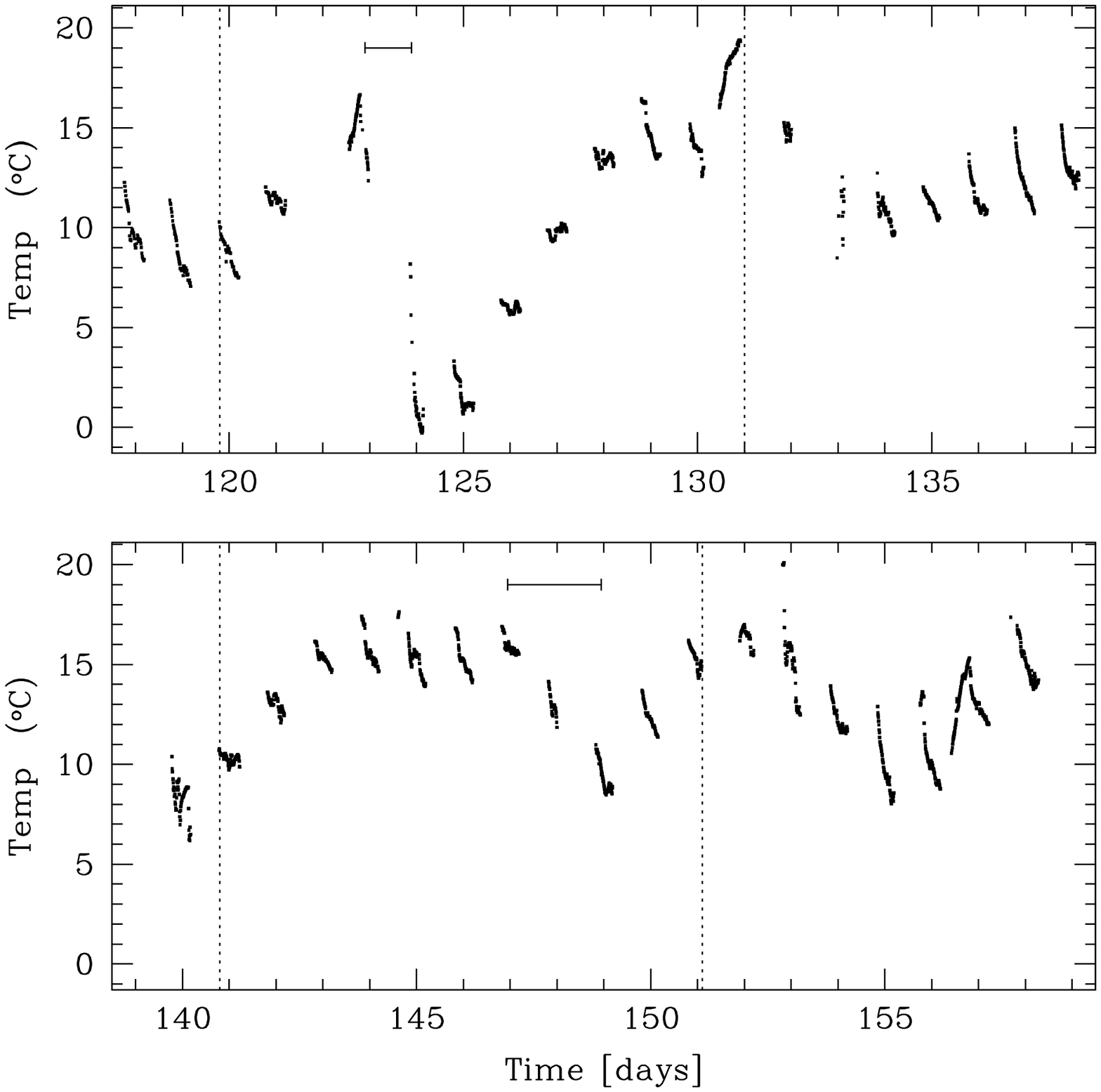}
	\figcaption{Temperatures recorded in the MMT telescope dome
between 27 April and 6 June 2001.  The bar in the upper panel shows an
extreme 17 $^{\circ}$C change in 24 hours.  The bar in the lower panel
shows a moderate 8 $^{\circ}$C change over 48 hours.  The dotted lines
show the time intervals we used for the extreme (upper panel) and
moderate (lower panel) boundary conditions.  Most of the results we
describe in the text refer to the moderate temperature conditions.  
The temperature gradients are a factor of two worse for the extreme
conditions.\\
\label{fig:temphist}}

\section{FIRST-ORDER CALCULATIONS}

We begin the heat transfer analysis with analytic calculations of
thermal capacitances and time constants.  We use these first-order
calculations to determine the level of detail needed in our models and
to check our finite difference calculations.

\subsection{Measurements of the Thermal Environment}

	Figure \ref{fig:temphist} shows MMT dome temperatures recorded
between 27 April and 6 June 2001.  We use the time periods between the
dotted lines in the upper and lower panels to represent extreme and
moderate environmental conditions, respectively.  In what follows, we
discuss the results appropriate for the moderate thermal environment
(ambient temperature changes by up to 8 $^{\circ}$C over 48 hours).  
We have also examined the results for the extreme thermal environment
(ambient temperature changes by up to 17 $^{\circ}$C in 24 hours); the
gradients and thermal offsets almost exactly double in this case.  The
large temperature changes in the extreme case are usually associated
with unsettled weather, so that a closed dome will likely be the
dominant concern.  In any case, it is easy to scale the results we
present for the extreme thermal environment.

\subsection{Thermal Capacitances}

The thermal capacitance is the energy stored in an object per unit
temperature, and is the product of the mass and the specific heat of
the object, $mC$.  The temperature change $\Delta T$ produced by a
heat input $q$ is: $\Delta T = {q \over {mC}} $ (see Equation
\ref{eqn:mcdt}). The thermal capacitance is therefore a useful tool
for understanding the relative thermal properties of instrument
components. Table \ref{tab:mc} lists the thermal capacitances of the
Binospec lens groups and their lens barrels.  We can immediately draw
two conclusions. First, because the aluminum lens barrels have much
smaller thermal capacitances than the lens groups, they will
experience greater temperature changes than the lens groups for a
given heat input. The lens barrels will therefore enforce radial
temperature gradients in the lenses.  We will need to model a lens
group with a number of radial slices (e.g.\ Figure
\ref{fig:lensgrp}).  Second, the second lens groups of both the
collimator and camera have approximately twice the thermal
capacitance of the neighboring lens groups and for a fixed heat input
they will experience smaller temperature changes than their
neighbors.  We will need to model the large lens groups with a number
of axial slices (see Figure \ref{fig:lensgrp}) to look for axial
temperature gradients.

\subsection{Thermal Time Constants} \label{sec:timeconstant}

Thermal time constants are useful characterizations of instrument
components.  An object out of thermal equilibrium with its environment
will have a temperature difference that exponentially decays to zero.
We calculate the thermal time constant associated with the exponential
decay assuming that the object is isothermal, and that the environment
has infinite thermal capacitance.  The thermal time constants for
conduction, convection, and radiation are:
	\begin{eqnarray}
	\tau_{cond} = & mC L / k A_x\label{eqn:taucond}\\
	\tau_{conv} = & mC / h A\\ 
	\tau_{rad} = & mC / h_r A\label{eqn:taurad}
	\end{eqnarray}

\includegraphics*[width=3.25in]{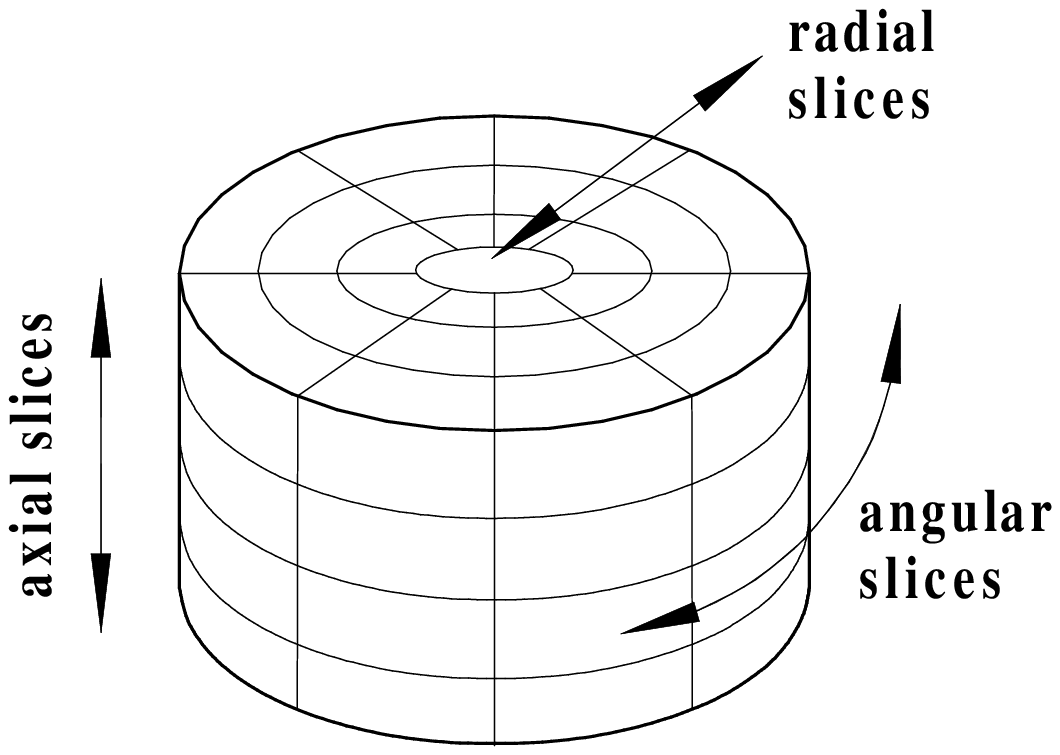}
	\figcaption{ \label{fig:lensgrp} Example of a Binospec lens
group model used in the detailed collimator model.  The radial and
axial slices are necessary to measure radial and axial temperature
gradients, respectively.  The angular slices are necessary to measure
asymmetric diametral gradients.\\}

\vskip 20pt

\noindent where $h_r = \epsilon \sigma (T + T_{env})(T^2
+ T_{env}^2) \simeq 4 \epsilon \sigma T_{av}^3$.  We use $h=2$ W
m$^{-2}$ K$^{-1}$ for still air in an enclosed cavity, and $h_r=5$ W
m$^{-2}$ K$^{-1}$ for temperatures near 0 $^\circ$C.

We model an idealized Binospec lens barrel, shown in Figure
\ref{fig:example2}, to illustrate thermal time constant calculations.
Our goal is to calculate how quickly heat conducts around a lens
barrel from a motor attached to the base of the barrel. We evaluate
the time constants in Equations \ref{eqn:taucond} - \ref{eqn:taurad}
for a \hbox{$m=20$ kg} aluminum barrel ($k=164$ W m$^{-1}$ K$^{-1}$,
$C=962$ J kg$^{-1}$ K$^{-1}$) with radius \hbox{0.25 m}, cross-section
\hbox{$A_x=0.005$ m$^2$}, and surface area \hbox{$A=0.4$ m$^2$}.  The
time constant for heat to conduct halfway around the lens barrel is
$\tau_{cond}=2.6$ hours.

We now consider the effects of convection and radiation on the
idealized lens barrel.  If its outer surface is exposed to the
environment, the lens barrel will equilibrate to the environment with
$\tau_{conv}=7$ hours and $\tau_{rad}=3$ hours.  Convective and
radiative heat transfer occur in parallel, and the summed radiative
and convective time constant is 2 hours.


\vskip 25pt

\includegraphics*[width=3.25in]{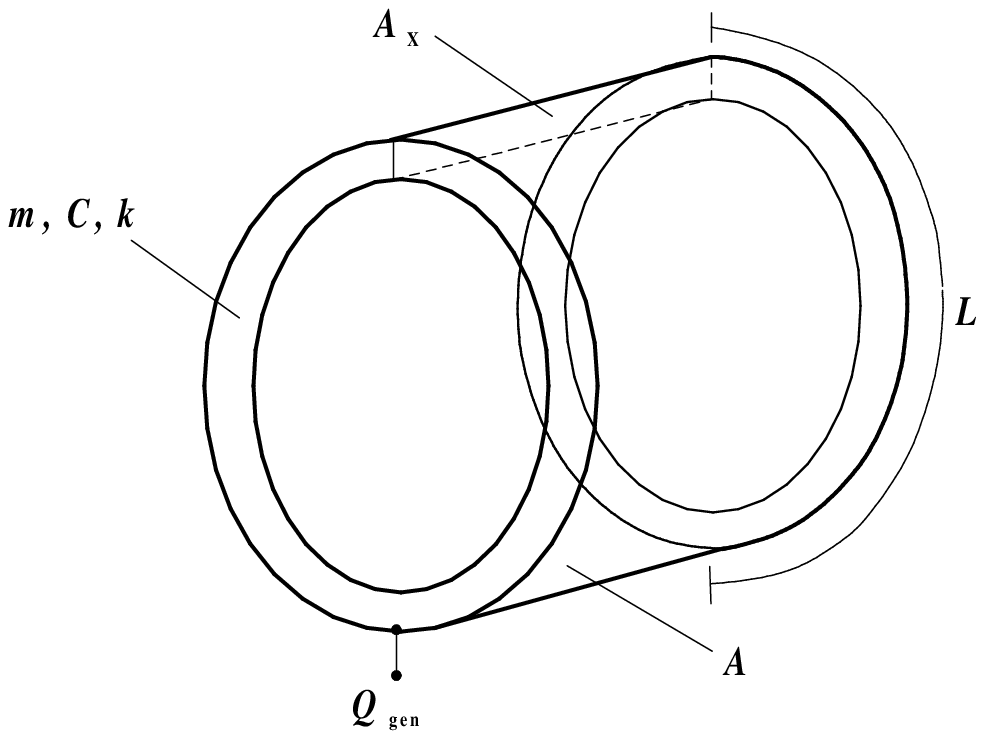}
\figcaption{ \label{fig:example2} Idealized Binospec lens barrel.\\}

If the interior of the lens barrel is filled with glass lenses
($k\simeq0.95$ W m$^{-1}$ K$^{-1}$, $C\simeq650$ J kg$^{-1}$
K$^{-1}$), the time constant for heat to conduct through the surface
area \hbox{$A=0.4$ m$^2$} to the lens center, \hbox{$L=0.25$ m}, is
$\tau_{cond}=18$ hours.  Thus, the lens barrel time constant is
$\sim$10 times shorter than the lenses it supports.  The difference
in time constants means that the lens barrel will act as an
approximately uniform temperature boundary condition for the lenses;
this allows us to use the lens barrel temperatures from our
low-resolution model as boundary conditions for our high resolution
models of the lens groups.  We use the detailed models, considering
exactly how the Binospec lenses are attached to the lens barrels, to
calculate the temperature gradients in the optics.

\section{FINITE DIFFERENCE THERMAL MODELS}

	We create finite difference thermal models of the instrument.
Each piece of the instrument, such as a lens or lens barrel, is broken
into isothermal lump mass elements.  Finite difference models for a
complicated system like the Binospec spectrograph and its components
require hundreds of nodes, with each node connected conductively,
convectively, and radiatively to dozens of other nodes. Writing out the
heat transfer equations for such a complicated system is difficult.  
We use the Thermal Analysis Kit III (TAK3), a professional thermal
analysis package written by K\&K Associates, to calculate the large
array of heat transfer equations. TAK3 has been previously used by
engineers at the Center for Astrophysics to thermally model the Chandra
X-ray Observatory.

	The heart of the TAK3 software is a general purpose finite
differencing thermal analyzer.  The masses, dimensions, surface
areas, and emittances of the elements, and the conductive,
convective, and radiative pathways between them are determined by the
user and typed into ASCII input files.  TAK3 allows the user to
easily modify the model assumptions and boundary conditions and to
reanalyze the system.

\subsection{Binospec Models}

	The lens barrels, mounts, and support struts are modeled with
single elements in the low-resolution model of the Binospec
spectrograph. Figure \ref{fig:ccnode} shows that each lens group in
the low-resolution model is modeled with a total of four elements,
dividing the group radially and axially.  This level of resolution is
sufficient to estimate center-to-edge radial gradients and
front-to-back axial temperature gradients in the lens groups.

	Radiation exchange is relatively difficult to handle unless a
complex instrument is modeled at low resolution.  View factors must
be calculated for every pair of elements that exchange radiation.
Convection is easier to handle because each compartment in the model
is assigned an air node that convectively couples the elements.  We
assume a convection coefficient of 2 W m$^{-2}$ K$^{-1}$ for still
air inside the instrument.  We use the average wind speed at the MMT,
$6\pm3$ \hbox{m s$^{-1}$} \citep{milone99}, to calculate the
convection coefficient exterior to the spectrograph.  Sequences of
temperature measurements recorded at the MMT are used as the model's
boundary condition.

	The high-resolution models of the collimator and camera optics,
by comparison, divide each lens group into multiple axial, radial, and
angular slices.  Figure \ref{fig:campic} illustrates the

\includegraphics*[width=3.25in]{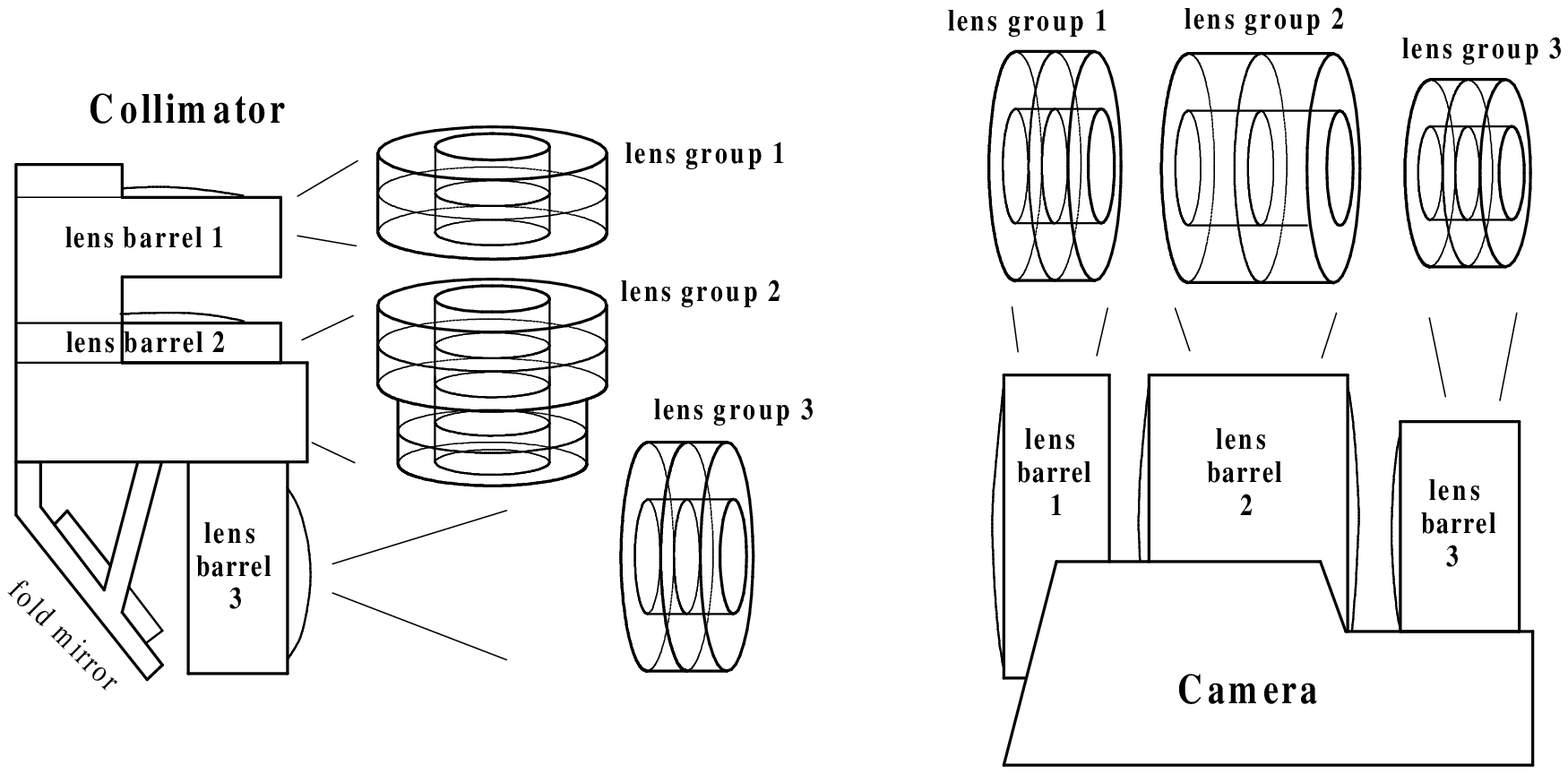}
\figcaption{ \label{fig:ccnode} Drawings of the collimator and
camera elements in the low resolution Binospec model.\\}

\includegraphics*[width=3.25in]{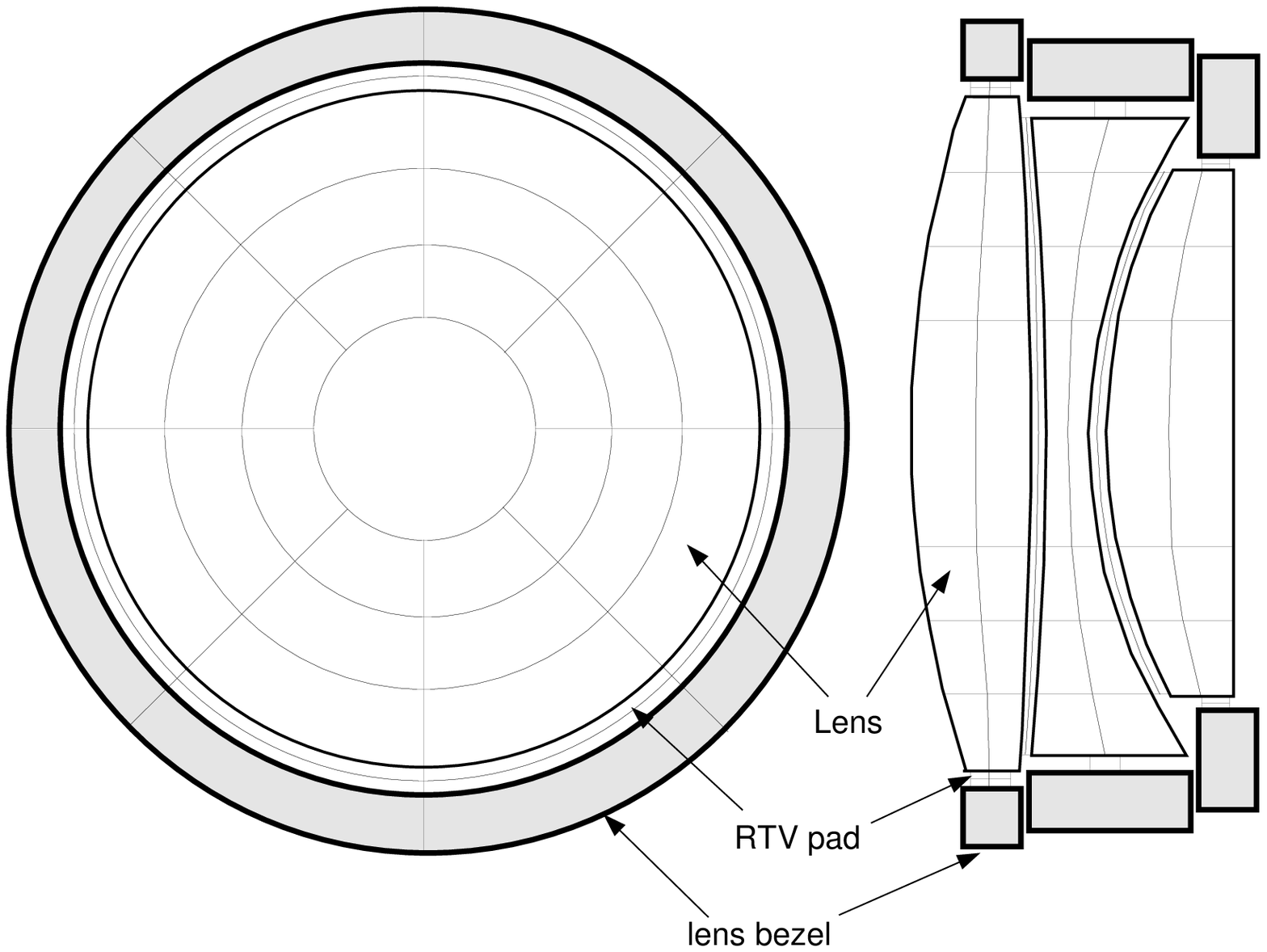}
\figcaption{ \label{fig:campic} Drawing of camera lens group 3 in the
detailed camera model.  Each optical element is divided into multiple
axial, radial, and angular slices.  Spaces between the lenses and the
lens bezels are exaggerated to show the coupling fluid layers and the
RTV bonds.\\}

\noindent lens group 3 portion of the detailed camera model.  
Spaces between the lenses, and the RTV layers connecting the lenses to
the aluminum bezels, are exaggerated for clarity.  The coupling fluid
that we use in the Binospec lens groups, Cargille Laser Liquid 5610, is
an excellent insulator ($k=0.147$ W m$^{-1}$ K$^{-1}$; B.\ Sutin 2002,
private communication) and so we model the coupling fluid layers with
multiple slices as well (see Figure \ref{fig:campic}).

	We find that the axial and radial temperature gradients
calculated for the lens groups generally agree to better than $0.02$
$^\circ$C between the low and high-resolution models.  However, the
high-resolution models reveal a new result: a diametral gradient in the
lens groups because the lens barrels are mounted to the optical bench
along one side. The amplitude of the diametral gradient is
approximately 25\% of the radial gradient.

\section{FINITE DIFFERENCE THERMAL MODEL RESULTS}

	The thermal models allow us to calculate heat flows,
temperature gradients, and thermal time constants.  Our basic picture
of heat transfer in Binospec comes from understanding how the
instrument responds to the telescope dome environment.  Because we
are interested in maintaining image quality in the spectrograph, our
main emphasis is on the collimator and camera lens groups and the
optical bench to which they are mounted.  In addition, we wish to
determine the thermal consequences of operating 

\includegraphics*[width=3.25in]{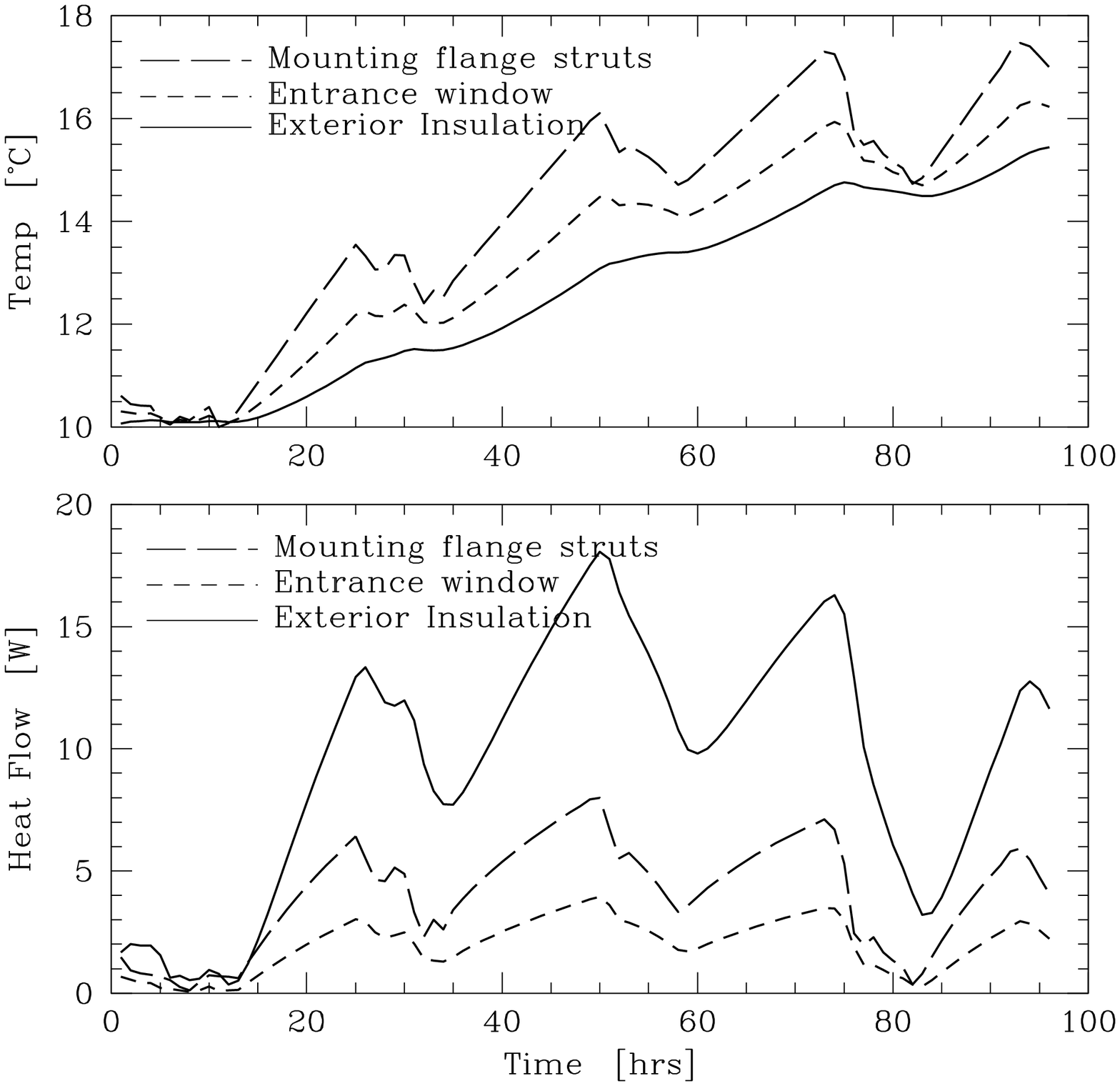}
	\figcaption{ \label{fig:qq} Binospec temperatures ({\it upper
panel}) and heat flows ({\it lower panel}) resulting from moderate MMT
temperature conditions.\\}

\noindent motors, as well as inserting new aperture masks, filters, and
gratings.  We are interested in how much insulation surrounding the
spectrograph is necessary.  We will use coefficients of thermal
expansion to calculate thermal deformations of the instrument structure
and to estimate the rate of thermally induced image drift on the
detector caused by thermal deflections of the optical bench.

\subsection{Heat Flow}

	Heat flow is the driving force in the thermal models.  The
greatest source of heat flow will dominate the thermal effects in the
instrument. Because the average heat flow with the environment must
be zero in the absence of internal heat sources, we consider {\it
peak} heat flows. Here we compare the peak heat flow from the
environment with the peak heat flow from operating motors, using the
low-resolution thermal model of Binospec.

	The peak heat exchange between the MMT telescope environment
and Binospec is $\sim$30 W.  Approximately 60\% of the heat flow passes
through the large surface area of the instrument; 40\% passes through
the mounting flange and the spectrograph entrance window.  Figure
\ref{fig:qq} plots the temperatures and heat flows through the exterior
insulation, mounting flange struts, and the entrance window for a
representative 100 hours of MMT operating conditions.  This result
assumes 75 mm thick urethane foam insulation surrounding the
spectrograph and moderate MMT dome temperatures that vary by up to 8
$^\circ$C over 48 hours.  Because the MMT primary mirror is actively
controlled to the ambient temperature with a forced air system, we
consider the telescope mounting flange to be at ambient temperatures at
all times. Table \ref{tab:qflow} summarizes the peak heat flows.

	The Binospec optical bench is attached to the mounting flange
with graphite epoxy struts (see Figure \ref{fig:binotop}). If the
conductivity (and heat flow) of the graphite epoxy struts were to
increase by a factor of three, then the time constant of the optical
bench is reduced from 37 to 27 hours and the temperature gradients in
the optics are increased by 10\%.  If the graphite epoxy struts were
thermally insulated on 1 cm thick Delrin pads, the heat flow is reduced
by a factor of four compared to the baseline model and the time
constant of the optical bench increases from 37 to 45 hours.

	The peak heat flow from internal motors when Binospec is used
for spectroscopy (changing an aperture mask, rotating the gratings,
and moving the guide probes every hour) or imaging (assuming a filter
change every 5 minutes) is $\sim$3 W, $\sim$10\% of the peak heat
flow from the environment.  This heat flow calculation assumes that
the motors are powered only when they are required to supply torque.
However, we must still look for local temperature gradients.  For
example, when imaging, the casings of filter motors may rise a few
$^\circ$C above the ambient temperature and induce temperature
gradients in collimator lens group 1.  We explore these issues in
subsection \ref{sec:heatsource}.

\subsection{Temperatures and Temperature Gradients}

	One of our primary concerns is that temperature gradients may
degrade the optical performance of the instrument.  Conveniently, the
primary output of the thermal models is temperatures for each element
as a function of time.  Here, we describe temperature variations in
the Binospec optics.

	Figure \ref{fig:colplot} illustrates the collimator lens group
temperatures from the low-resolution Binospec model. The upper panel
shows the cores of the three collimator lens groups responding to the
moderate external temperature fluctuations.  Lens group 2 has
approximately twice the thermal capacitance of the other two lens
groups and lags them by six hours; the resulting worst-case temperature
differences of $\sim$0.5 $^\circ$C are shown in the lower panel.  
Figure \ref{fig:camplot} shows the comparable results for the camera
lens groups.

	Figure \ref{fig:colplot2} plots the worst case radial
temperature gradient in the Binospec optics, located in collimator lens

\includegraphics*[width=3.25in]{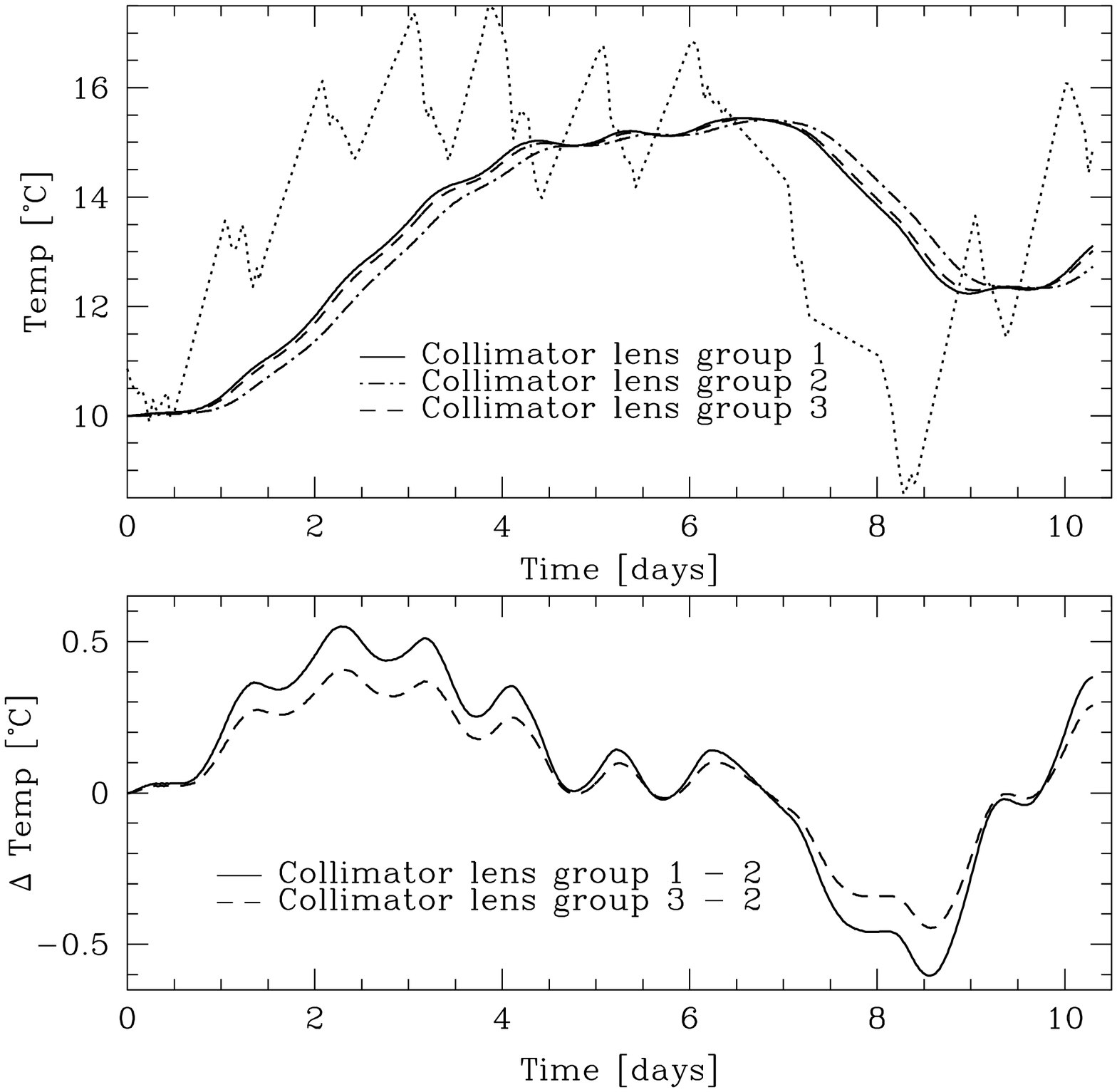}
	\figcaption{ \label{fig:colplot} Binospec collimator lens group
temperatures ({\it upper panel}) and temperature differences between
lens groups ({\it lower panel}).  The dotted line shows the moderate
MMT dome temperatures used as the boundary condition for the model.\\}

\includegraphics*[width=3.25in]{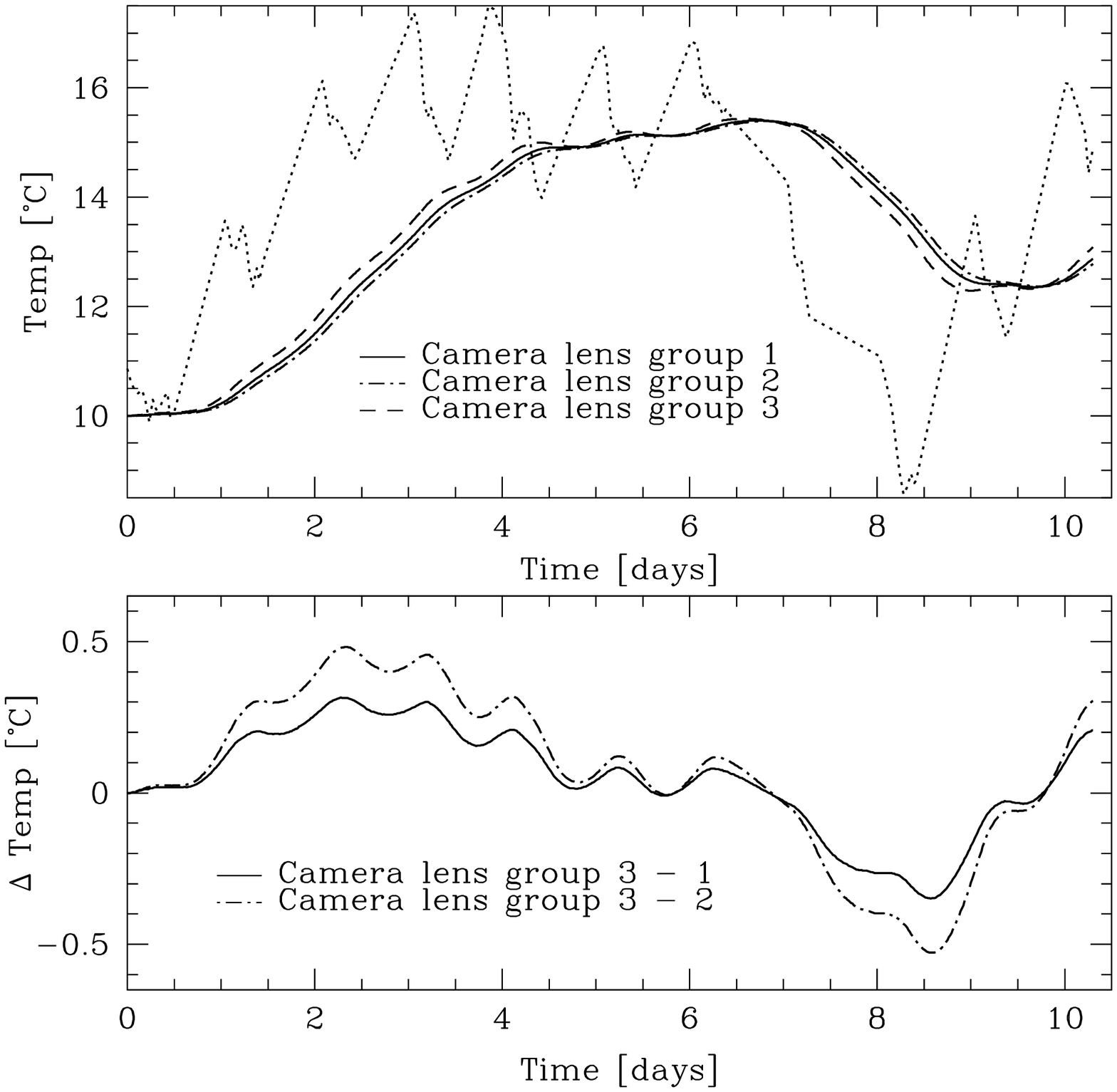}
	\figcaption{ \label{fig:camplot} Binospec camera lens group
temperatures ({\it upper panel}) and temperature differences between
lens groups ({\it lower panel}).  The dotted line shows the moderate
MMT dome temperatures used as the boundary condition for the model.\\}

\noindent group 2.  The detailed collimator model shows that this
radial temperature gradient peaks at 0.11 $^\circ$C under moderate
conditions. Similarly, Figure \ref{fig:camplot2} shows the worst case
axial temperature gradient, located in camera lens group 3.  The
detailed camera model shows that this axial gradient peaks at 0.14
$^\circ$C. \cite{epps02} have shown that 0.2 $^\circ$C gradients
negligibly affect image quality and image scale.

	We have created additional models using different environmental
boundary conditions and different designs.  We have compared the
results with the baseline case shown 

\includegraphics*[width=3.25in]{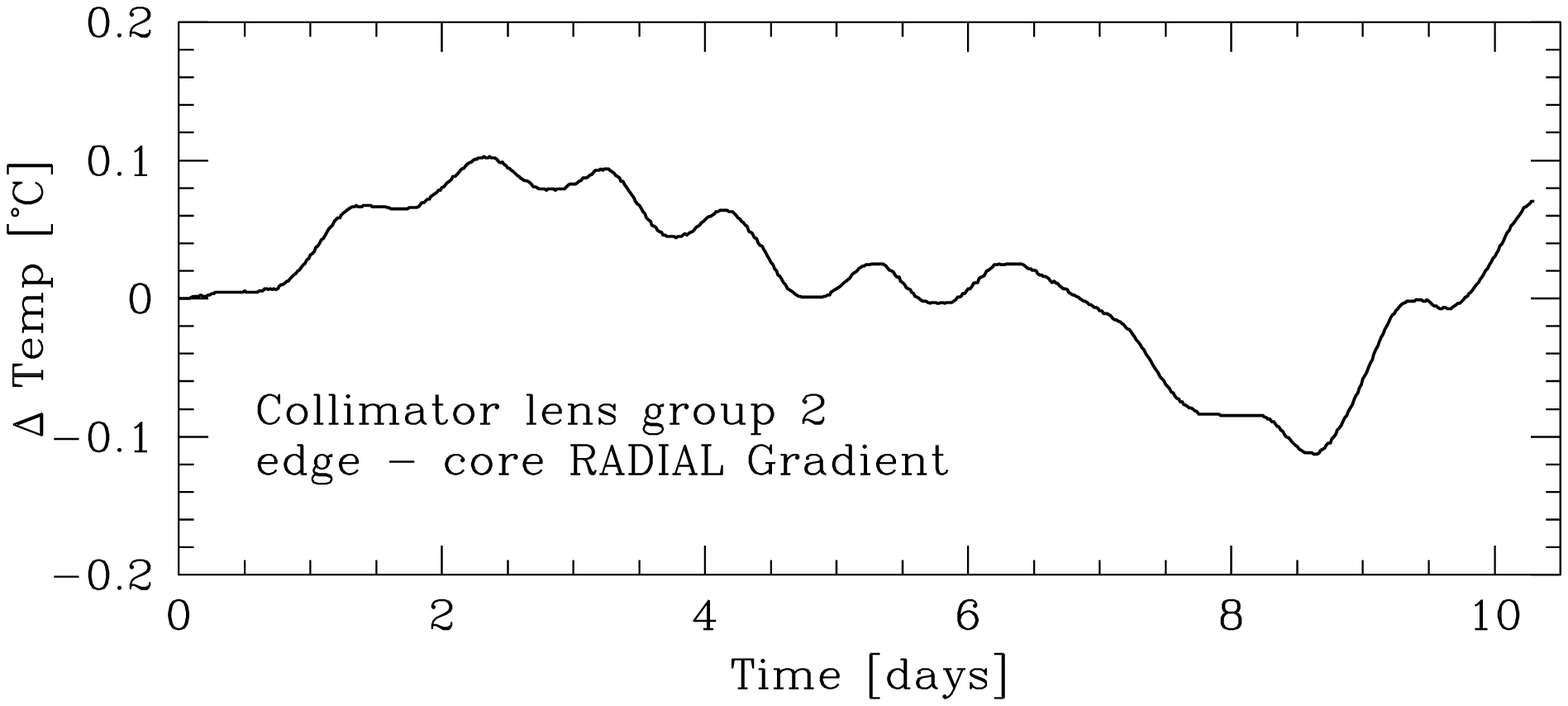}
	\figcaption{ \label{fig:colplot2} Worst case radial temperature
gradient, Binospec collimator lens group 2.\\}

\includegraphics*[width=3.25in]{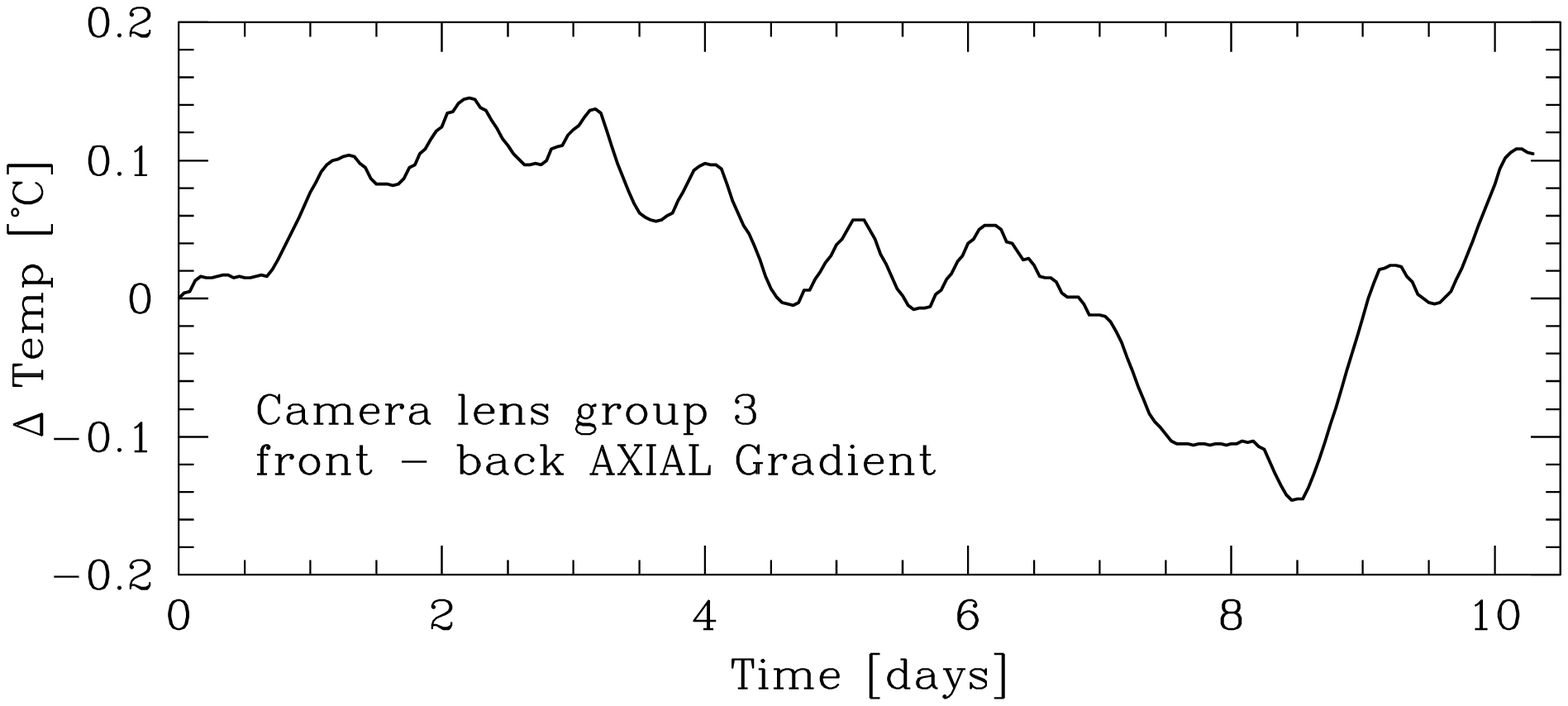}
	\figcaption{ \label{fig:camplot2} Worst case axial temperature
gradient, Binospec camera lens group 3.\\}

\noindent in Figures \ref{fig:colplot} - \ref{fig:camplot2}.  For
example, we find that removing the spectrograph entrance window will
result in $\sim$0.2 $^\circ$C ($\sim$100\% larger) temperature
gradients in the optics and a 26 hour (30\% shorter) optical bench
thermal time constant.

\subsection{Internal Heat Sources} \label{sec:heatsource}


Internal heat sources generate heat flow that cause temperature
gradients and temperature transients.  Binospec uses motors to move
guide probes, and to change aperture masks, filters, and gratings.
The motors used to change filters may be operated frequently when
Binospec is used for imaging.  The Binospec filter changer sits
directly above collimator lens group 1 (see Figure \ref{fig:binotop})
and uses four motors, each of which may generate 16 W for 10 s during
a filter change.  If the filter changer is operated once every five
minutes for ten hours, Equation \ref{eqn:mcdt} tells us the entire
spectrograph will be heated by $\sim$0.05 $^\circ$C.  Though this is a
small temperature change, we must consider the effect of larger local
temperature gradients.

\includegraphics*[width=3.25in]{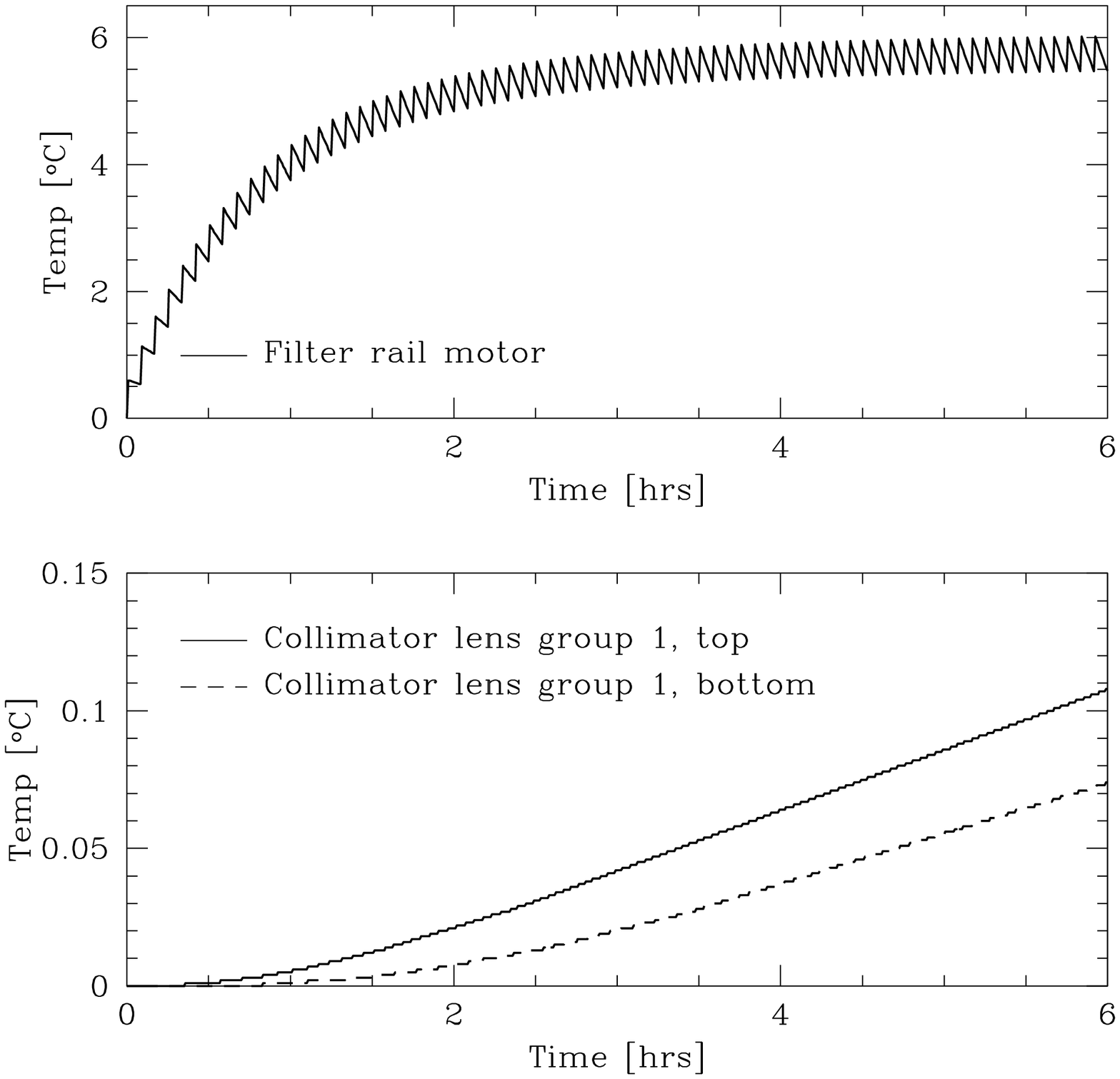}
	\figcaption{ \label{fig:filmotor} Temperature of a Binospec
filter changer motor ({\it top panel}) generating 16 W for 10 s every
five minutes, and the resulting temperature of the nearby collimator
lens group 1 ({\it lower panel}).\\}

Figure \ref{fig:filmotor} plots the time dependent temperatures of a
filter changer motor and the top and bottom of collimator lens group
1.  The motor temperature rises 6 $^\circ$C with a time constant of
$\sim$1 hour.  After a few hours of operation, the heat flow from the
motors causes a 0.03 $^\circ$C top-to-bottom axial gradient in
collimator lens group 1.  Collimator lens group 2, on the other hand,
experiences an uniform rise in temperature with no temperature
gradient.  We conclude that using the Binospec filter motors every
five minutes for a few hours will cause {\it local} temperature
gradients in the optics smaller than those introduced by the telescope
environment (e.g.~Figure \ref{fig:colplot2}).

\subsection{Time Constants}

Binospec's internal temperature gradients would be minimized if all of
its internal parts could be designed to have short time constants.
Unfortunately, the large thermal capacitances of the optical elements
makes this

\includegraphics*[width=3.25in]{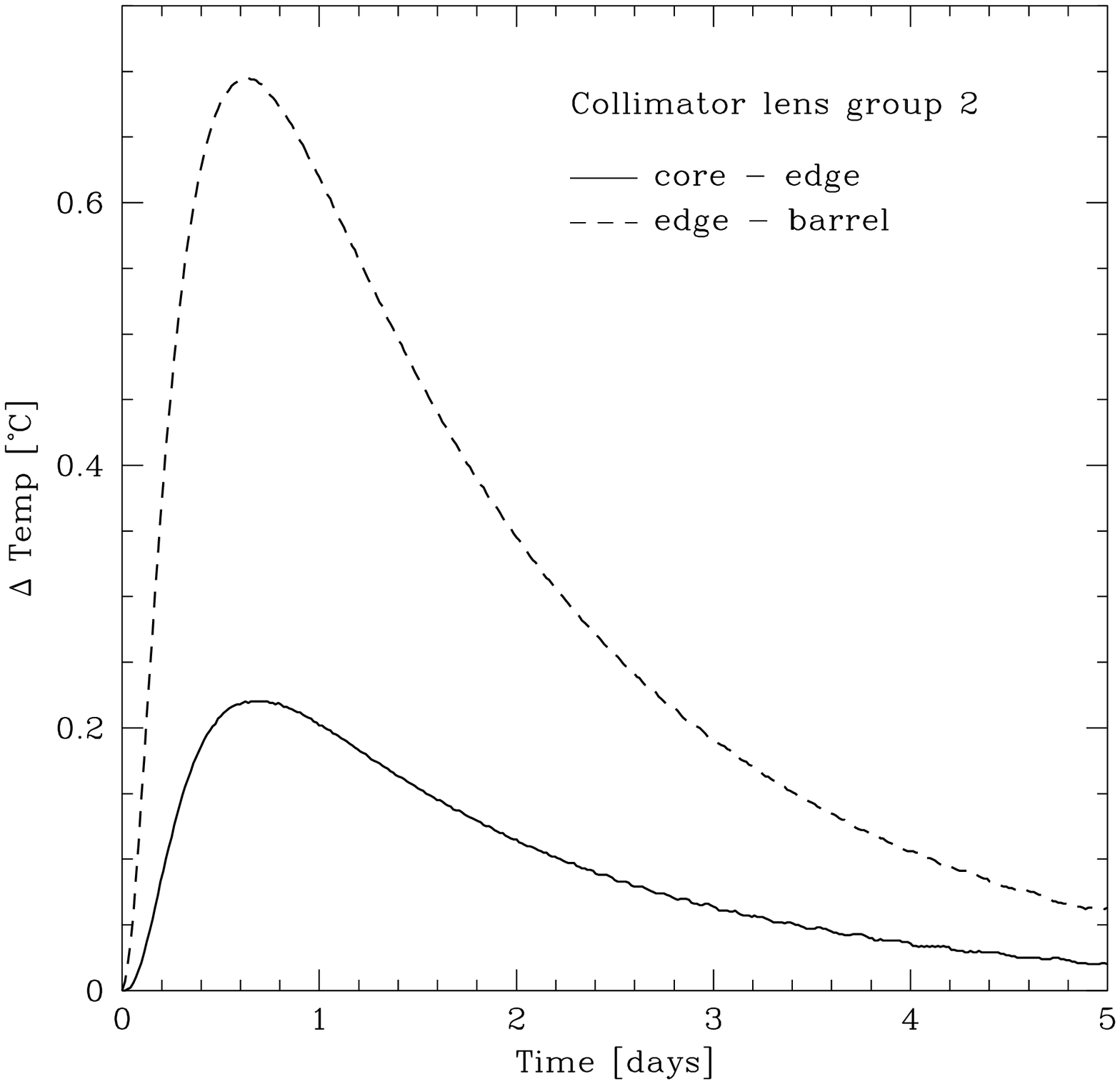}
	\figcaption{ \label{fig:tauplot} Temperature differences
between the Binospec collimator lens group 2 elements, assuming the
model starts 10 $^\circ$C above the environment and then
equilibrates.\\}

\noindent impractical.  The best we can do is to heavily insulate
Binospec from the environment to slow heat exchange.

The thermal time constants presented here are analogous to those in
Equations \ref{eqn:taucond} to \ref{eqn:taurad}.  The thermal time
constants quantify the response time of components in the thermal
model, and are calculated by starting the thermal model at an uniform
temperature 10 $^\circ$C above the environment and finding the
$e$-folding equilibration time for each component.  Large differences
between time constants of adjacent components are undesirable because
they lead to large temperature gradients.

Thermal time constants for the Binospec collimator lens group 2, for
example, are summarized in Table \ref{tab:binotau}.  The collimator
lens group 2 has an $\sim48$ hour thermal time constant in the well
insulated Binospec design. There is a small difference in time
constant from the center to the edge of the collimator lens group and
a corresponding $\sim$0.2 $^\circ$C temperature gradient for a 10
$^\circ$C temperature change (Figure \ref{fig:tauplot}). The lens
barrel, however, has a 10\% shorter time constant than the lens group
and a corresponding $\sim$0.7 $^\circ$C temperature gradient between
the barrel and the lens group.  This difference in thermal time
constant is in agreement with the time constant calculations in section
\ref{sec:timeconstant}.

\subsection{Inserting Aperture Masks and Filters}

Changing aperture masks in a multiobject spectrograph like Binospec is
a daily task.  Opening the instrument to change aperture masks,
filters, or gratings may expose the instrument to a troublesome
thermal shock.  New components placed in the instrument will most
likely be at a different temperature than the well insulated
instrument and will generate heat flows and temperature gradients
inside the instrument.

The aperture masks and filters are located in a compartment that
includes the first collimator lens group and the top side of the
optical bench (see Figure \ref{fig:binotop}).  We assume that a set of
ten 0.5 kg aperture masks or a set of twelve 

\includegraphics*[width=3.25in]{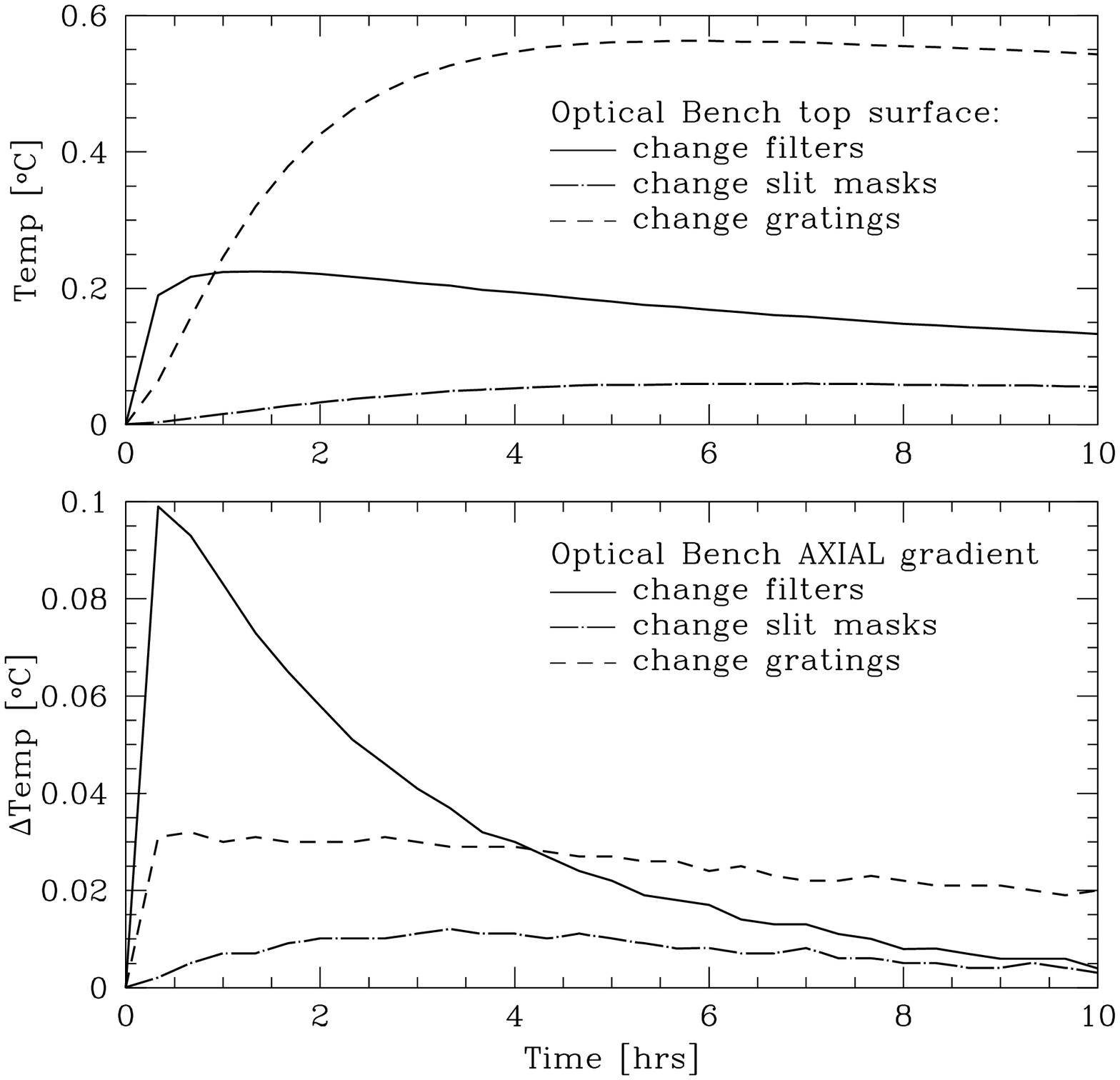}
\figcaption{ \label{fig:optbpap} Temperature changes ({\it upper
panel}) and temperature gradients ({\it lower panel}) in the optical
bench resulting from inserting new filters, aperture masks, and
gratings.  The models assume that the new components and the
surrounding air are 10 $^\circ$C above the ambient instrument
temperature.  These temperature changes and gradients add to the
underlying temperature changes and gradients in the baseline
model.\\}

\noindent 1.2 kg filters will be changed at a time, and that the masks
and filters, as well as the air introduced into the compartment are an
extreme 10 $^\circ$C hotter than the instrument.  Figure
\ref{fig:optbpap} shows the resulting temperature changes and axial
temperature gradients in the optical bench, where temperature gradients
may cause significant thermal deflection.  These temperature changes
and gradients add to the underlying temperature changes and gradients
in the baseline model.

For the case of inserting new aperture masks, the optical bench is
heated 0.05 $^\circ$C and experiences a 0.01 $^\circ$C axial gradient
with a time constant of three hours.  For the case of inserting new
filters, the optical bench is heated 0.1 $^\circ$C and experiences a
peak 0.1 $^\circ$C axial gradient with a time constant of four hours.
The first collimator lens group, on the other hand, experiences a
nearly uniform $\sim$0.1 $^\circ$C rise with a time constant of four
hours.  These temperature changes scale with the initial 10 $^\circ$C
temperature difference.  We conclude that the temperature gradients
induced by inserting aperture masks or filters are small and will
equilibrate with an acceptable time constant.

The six diffraction gratings have ten times the thermal capacitance of
the aperture masks and filters, and so changing gratings is more of an
issue.  The six 9 kg gratings plus their 7 kg grating holders will
heat the entire instrument by 0.5 $^\circ$C.  Secondly, the gratings
have a poor conduction path to the rest of the instrument, and their
equilibration time constant is long.  Finally, the gratings directly
view the collimator and camera lens groups.  For an initial 10
$^\circ$C temperature difference, the gratings will induce a 0.1
$^\circ$C axial gradient through collimator lens group 3 and camera
lens group 1.

\clearpage 

\subsection{Thermal Stress}

Temperature gradients cause thermal stress in the optical elements.
In this section we estimate the thermal stresses in a simplified
Binospec lens.  We approximate a lens by a cylindrical
disk, and use the following stress equations from \citet{roark}.
	The total stress caused by an uniform axial temperature gradient,
$\Delta T_z$, in a disk clamped around its edge is given by:
	\begin{equation}
	  S = 1/2 ~\alpha E \Delta T_z /(1-\nu),
	\end{equation} 
where $S$ is the stress, $\alpha$ is the coefficient of thermal
expansion, $E$ is the Young's modulus, and $\nu$ is the Poisson's
ratio of the material.  The thermal stress caused by a 0.1 $^\circ$C
axial gradient in a typical glass lens ($\alpha \simeq 8\times10^{-6}$
$^\circ {\rm C}^{-1},~E \simeq70\times10^9~{\rm Pa},~\nu \simeq 0.24$)
is 5 psi.  The thermal stress caused by a 0.1 $^\circ$C axial gradient
in a CaF$_2$ lens ($\alpha=19\times10^{-6}$ $^\circ {\rm C}^{-1},~E
=76\times10^9~{\rm Pa},~\nu=0.26$) is 14 psi.

The stress in a disk caused by an uniform radial temperature
gradient, $\Delta T_r$, is given by:
	\begin{eqnarray}
S_r = 1/3~\alpha E \Delta T_r (r_o -r)/r_o{\rm ~and}\\
S_t = 1/3~\alpha E \Delta T_r (r_o -2r)/r_o,
	\end{eqnarray} where $S_r$ and $S_t$ are the stresses in
the radial and tangential directions, respectively, and $r_o$ is the
outer radius. Note that both the radial and tangential stresses peak
in the lens center ($r=0$).  If we sum the radial and tangential
stress in the lens center, the total stress caused by a 0.1 $^\circ$C
radial gradient is 5 psi for a typical glass lens and 14 psi for
CaF$_2$.  These stress levels are insignificant.

\subsection{Thermal Deflections}

One of our design goals is to minimize image drift at the detector.
Image drift degrades image quality and the wavelength calibration
during a long spectroscopic exposure.  The thermal stability of the
Binospec optical bench is a key issue in understanding image drifts.
We can simply estimate the thermal deflection of the Binospec optical
bench by modeling the bench as an unconstrained circular disk.  The
disk geometry is a good description of the Binospec optical bench and
the formulas can be taken directly from \citet{roark}.  The radius of
curvature $R$ of a disk with a uniform, top-to-bottom temperature
difference $\Delta T_z$ is: 
\begin{equation} 
R =\frac{t}{\alpha \Delta T_z}, 
\end{equation} where $\alpha$ is the coefficient of thermal
expansion and $t$ is the thickness of the disk.  The thermal
deflection $\theta$ caused by the axial temperature difference is:
\begin{equation} 
\label{theta} 
\tan \theta =\frac{r/R}{\sqrt{1-(r/R)^2}}.  
\end{equation}

Binospec has a cylindrical aluminum optical bench that is 2.14 m in
diameter and 0.15 m thick.  In our thermal models, the optical bench
experiences up to $\pm0.1$ $^\circ$C axial temperature gradients.  The
resulting tilt at the edge of the optical bench is $\pm3.5$ arcsec. We
calculate the same tilt when we make a detailed finite element model of
the optical bench.  The optical bench responds to external temperatures
that vary with a time scale of twelve hours, so the rate of deflection at
the edge of the optical bench 

\includegraphics*[width=3.25in]{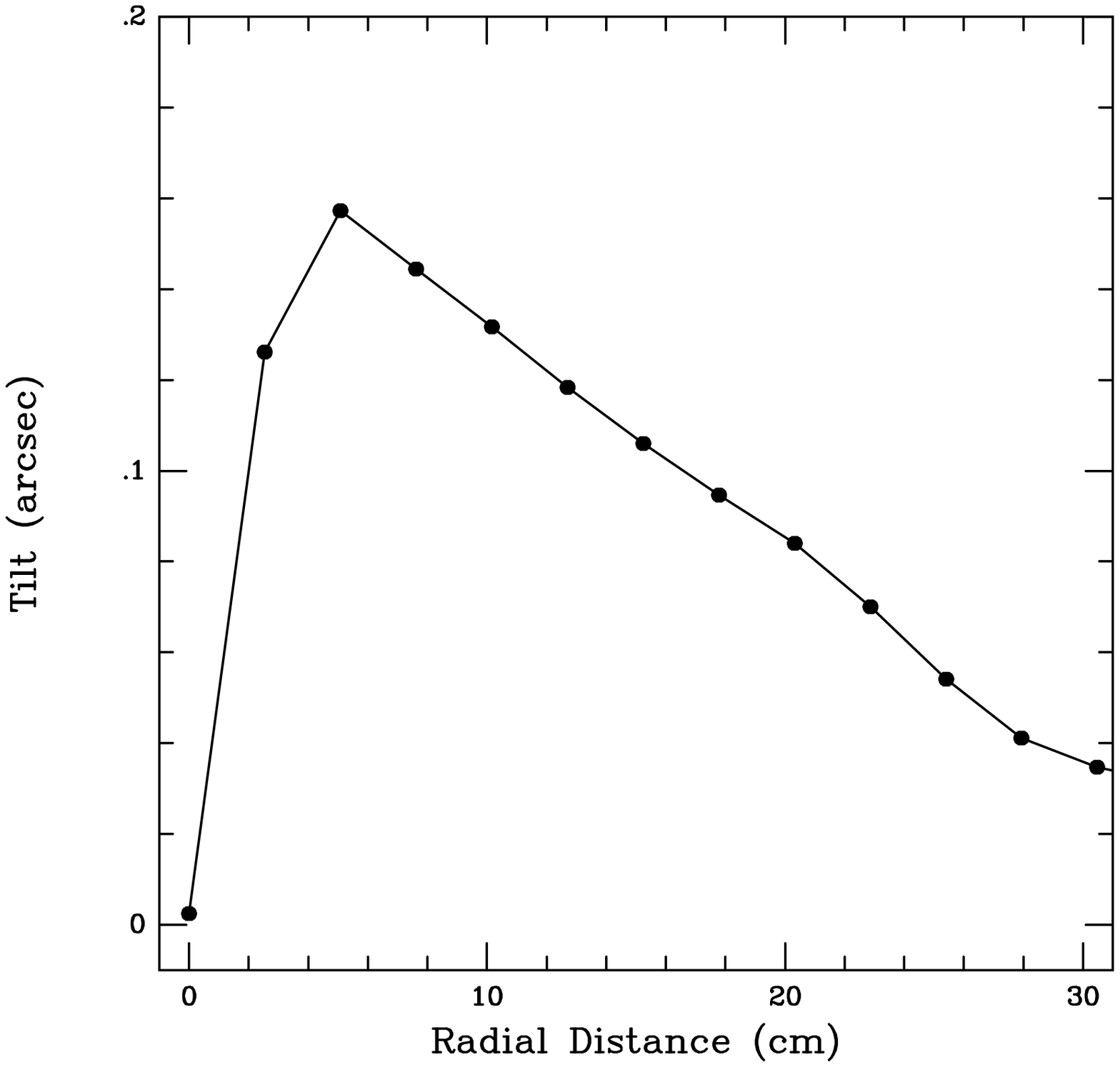}
	\figcaption{ \label{fig:spot} Deflection caused by a
0.1$^\circ$C hot spot at the center of the Binospec optical bench.\\}

\noindent is $3.5 / 12 = 0.3$ arcsec hr$^{-1}$.  If we
insert new filters into Binospec, the time constant is four hours and the
rate of deflection at the edge of the optical bench is 0.9 arcsec
hr$^{-1}$.

We use the ZEMAX optical design software (published by Focus Software,
Inc.) to estimate image shifts at the detector due to deflection of
various optical elements mounted on the optical bench.  The grating is
by far the most sensitive optical element: a 4 arcsec deflection of
the grating leads to a 1 pixel shift on the detector.  The grating
experiences 2/3 of the total optical bench deflection, so that
the image drift due to external temperature changes is an acceptable
0.05 pixel hr$^{-1}$.  We conclude that thermal deflection of the
optical bench by external temperature changes will not significantly
degrade an hour long spectroscopic observation.  The corresponding
result for image drifts following the insertion of new filters is 0.15
pixel hr$^{-1}$, so we will want to minimize filter changes.

More extreme temperature gradients may be created by internal heat
sources, such as a hot motor mounted on the optical bench.  We turn to
mechanical finite element modeling to assess the deflection caused by
a hot spot on the optical bench.  We find that a 0.1 $^\circ$C hot
spot, with a time constant of 1.5 hours, will cause a local 0.2 arcsec
deflection of the Binospec optical bench.  A 16 W motor mounted to the
optical bench on 6 mm thick Delrin standoffs will produce a 0.1
$^\circ$C hotspot if powered with a 3\% duty cycle.  Figure
\ref{fig:spot} shows the deflection as a function of distance from the
spot. If the grating mount were located next to this hot spot, the
resulting image drift rate would be 0.03 pixel hr$^{-1}$.  This is
similar to the drift rate caused by the 0.1 $^\circ$C gradient across
the optical bench from external temperature changes, and suggests
that we want to be careful how we mount motors near the grating.
Thermal standoffs are effective in reducing the magnitude of hot
spots.

\section{CONCLUSIONS}

We have presented the thermal analysis techniques that we have used to
study the thermal behavior of Binospec, a wide-field, multiobject
optical spectrograph that uses high performance refractive optics.  We
begin our analysis by calculating thermal time constants and thermal
capacitances to understand the scale of temperature variations in the
instrument and to determine what areas of the instrument require
detailed modeling.  We then generate a low-resolution finite difference
thermal model of the entire spectrograph, and high-resolution models of
the collimator and camera optics, optical bench, and filter changer.  
We use temperature data recorded at the MMT to set the boundary
conditions for the models.  Conduction, convection, and radiation are
all important modes of heat transfer in the instrument.

In normal operation, we find that the Binospec lens groups will
experience up to $\sim$0.14 $^\circ$C axial and radial temperature
gradients and that lens groups of different mass will have temperatures
that differ by $\sim$0.5 $^\circ$C. These temperature gradients are
driven by heat exchange with the environment; internal heat sources are
a minor contribution to the total heat flow.  Under extreme conditions
at the MMT, we find that these gradients and offsets double.  
\cite{epps02} have shown that these temperature gradients and
temperature offsets in the optics have a negligible effect on image
quality and image scale, even under the extreme conditions.

If the spectrograph is exposed to an environment at a 10 $^\circ$C
different temperature in the process of changing aperture masks or
filters, we find the optics will experience $<0.1$ $^\circ$C
temperature gradients, comparable to those experienced in normal
operation without opening the spectrograph.  The thermal stresses
associated with these gradients are negligible.

Our calculations show that the optical bench will experience a peak 0.1
$^\circ$C axial gradient in normal operation from external temperature
changes, and an additional 0.1 $^\circ$C axial gradient when the
spectrograph is opened to change filters.  The time constants for these
gradients are twelve hours and four hours, respectively.  The resulting
time dependent deformations of the optical bench will tilt the grating
mount and cause an 0.05 and 0.15 pixel hr$^{-1}$ image drifts at the
detector, respectively.  Overall, we conclude that the current Binospec
design has acceptable thermal properties.

Further thermal modeling will be helpful as we proceed to the detailed
design of Binospec.  Thermally isolating the optical bench from the
telescope mounting flange will reduce the conductive heat flow into
the spectrograph and reduce the temperature gradients in the optical
bench. Providing (passive) ventilation around the optical bench will
allow convection to equilibrate the temperature of the spectrograph
interior.  Mounting motors on thermal stand-offs will prevent local
hot spots in the structure and allow radiation and convection to
distribute heat around the spectrograph.  Radiation shields on the
motors may also be helpful to prevent motors from directly affecting
sensitive optical surfaces.  The spectrograph electronics will be
mounted in a temperature controlled enclosure that is insulated from
the spectrograph interior.

\acknowledgements	

We thank Bob Fata and Jack Barberis for their important contributions
to this project and Henry Bergner for providing the finite element
model of the optical bench.  We thank Brian Sutin for pointing out that
the lens coupling fluid may significantly affect temperature gradients
and time constants in the lens groups.  Steve West kindly supplied the
MMT dome temperature data.

\bibliographystyle{apj}
\bibliography{References}

\begin{thebibliography}{8}
\expandafter\ifx\csname natexlab\endcsname\relax\def\natexlab#1{#1}\fi

\bibitem[{{Epps}(1998)}]{epps98}
{Epps}, H.~W. 1998, in Proc. SPIE, Vol. 3355, {Optical Astronomical
  Instrumentation}, ed. S.~D'Odorico, 111--128

\bibitem[{{Epps} \& {Fabricant}(2002)}]{epps02}
{Epps}, H.~W. \& {Fabricant}, D.~G. 2002, accepted by PASP

\bibitem[{{Fabricant} {et~al.}(1998){Fabricant}, {Fata}, \&
  {Epps}}]{fabricant98}
{Fabricant}, D.~G., {Fata}, R.~G., \& {Epps}, H.~W. 1998, in Proc. SPIE, Vol.
  3355, {Optical Astronomical Instrumentation}, ed. S.~D'Odorico, 232--241

\bibitem[{{Fata} \& {Fabricant}(1993)}]{fata93}
{Fata}, R.~G. \& {Fabricant}, D.~G. 1993, in Proc. SPIE, Vol. 1998,
  Optomechanical Design, ed. D.~Vukobratovich, P.~R. Yoder, \& V.~L. Genberg,
  32--38

\bibitem[{{Fata} \& {Fabricant}(1998)}]{fata98}
{Fata}, R.~G. \& {Fabricant}, D.~G. 1998, in Proc. SPIE, Vol. 3355, {Optical
  Astronomical Instrumentation}, ed. S.~D'Odorico, 275--284

\bibitem[{{Incropera} \& {Dewitt}(1996)}]{incropera}
{Incropera}, F.~P. \& {Dewitt}, D.~P. 1996, Fundamentals of Heat Transfer (New
  York: Wiley, 1996, 4th ed.)

\bibitem[{{Milone} {et~al.}(1999){Milone}, {Heller}, \& {McAfee}}]{milone99}
{Milone}, A.~A.~E., {Heller}, C., \& {McAfee}, J. 1999, MMT Technical
  Memorandum 99-1

\bibitem[{{Roark} \& {Young}(1989)}]{roark}
{Roark}, R.~J. \& {Young}, W.~C. 1989, {Roark's Formulas for Stress and Strain}
  (New York: McGraw-Hill, 1989, 6th ed.)

\end{thebibliography}


\begin{deluxetable}{clc}	
\tablewidth{0pt}
\tablecaption{Symbols\label{tab:symbols}}
\tablehead{ \colhead{Symbol} & \colhead{Description} & \colhead{Units} }
	\startdata
$q$	& heat & J \\
$Q$	& heat transfer rate & W \\
$T$	& temperature & K \\
$k$	& conductivity & W m$^{-1}$ K$^{-1}$ \\
$h$	& convection coefficient & W m$^{-2}$ K$^{-1}$ \\
$\sigma$ & Stefan-Boltzmann constant & W m$^{-2}$ K$^{-4}$ \\
$L$	& length & m \\
$A$	& surface area & m$^2$ \\
$A_x$	& cross-sectional area & m$^2$ \\
$m$	& mass & kg \\
$C$	& specific heat (at constant pressure) & J kg$^{-1}$ K$^{-1}$ \\
$\alpha$	& coefficient of thermal expansion & m m$^{-1}$ K$^{-1}$ \\
$S$		& stress & Pa \\
$E$		& Young's modulus & Pa \\
$\nu$		& Poisson's ratio & \\
$\epsilon$	& emittance & \\
$F_{ij}$	& view factor & \\
\enddata
\end{deluxetable}

\begin{deluxetable}{rcc}	
\tablewidth{0pt}
\tablecolumns{3}
\tablecaption{Binospec Thermal Capacitances\label{tab:mc}}
\tablehead{
  \colhead{Object} & \colhead{lens group} & \colhead{lens barrel} \\
  \colhead{} & \colhead{$mC$ (J/K)} & \colhead{$mC$ (J/K)} }
	\startdata
Collimator & & \\
lens group 1 & 15700 & ~5900\\
lens group 2 & 26100 & 16200\\
lens group 3 & 12400 & ~6600\\
Camera & & \\
lens group 1 & 20800 & ~3200\\ 
lens group 2 & 47000 & 7600\\
lens group 3 & 13000 & ~4100\\
	\enddata
\end{deluxetable}

\begin{deluxetable}{lc}		
\tablewidth{0pt}
\tablecolumns{2}
\tablecaption{Peak Heat Exchange with Environment\label{tab:qflow}}
\tablehead{	\colhead{Component} & \colhead{$Q$} \\
		\colhead{} & \colhead{(Watts)} }
	\startdata
Exterior insulation	& 18 \\
Mounting flange struts	& 8 \\
Entrance window		& 4 \\
			& == \\
Total:			& 30 \\
	\enddata
\end{deluxetable}

\begin{deluxetable}{cc} 	
\tablewidth{0pt}
\tablecolumns{2}
\tablecaption{Binospec Time Constants\label{tab:binotau}}
\tablehead{	\colhead{Collimator lens group 2} & \colhead{$\tau$} \\
		\colhead{} & \colhead{(hrs)} }
	\startdata
barrel	& 43 \\
edge	& 47 \\
core	& 48 \\
	\enddata
\end{deluxetable}


\end{document}